\documentclass[11pt]{article}

\usepackage{setspace} 
\usepackage[utf8]{inputenc}
\UseRawInputEncoding

\usepackage[margin=1in]{geometry}  	 

\usepackage{pdfpages} 

\usepackage{hyperref} 

\usepackage[toc,title,page]{appendix} 

\usepackage[nottoc]{tocbibind} 

\usepackage{amsthm}
\newtheorem{theorem}{Theorem}[section] 

\usepackage{caption}
\usepackage{subcaption}

\usepackage{booktabs}
\usepackage{multirow}
\usepackage{graphicx}

\usepackage{amsmath}
\usepackage{amssymb}
\DeclareMathOperator*{\argmax}{argmax}
\DeclareMathOperator*{\argmin}{argmin}

\usepackage[ruled,noend]{algorithm2e}

\usepackage{authblk}

\makeatletter
\def\@seccntformat#1{\@ifundefined{#1@cntformat}%
   {\csname the#1\endcsname\quad}  
   {\csname #1@cntformat\endcsname}
}
\let\oldappendix\appendix 
\renewcommand\appendix{%
    \oldappendix
    \newcommand{\section@cntformat}{\appendixname~\thesection\quad}
}
\makeatother

\begin{document}

\title{\textbf{Tree-based methods for estimating heterogeneous model performance and  model combining} \vspace*{0.3in} }

\author[1]{Ruotao Zhang}
\author[1]{Constantine Gatsonis}
\author[1]{Jon Steingrimsson}

\affil[1]{Department of Biostatistics, School of Public Health, Brown University, Providence, RI}

\maketitle{}

\noindent \title{\textbf{Abstract:}} Model performance is frequently reported only for the overall population under consideration. However, due to heterogeneity, overall performance measures often do not accurately represent model performance within specific subgroups. We develop tree-based methods for the data-driven identification of subgroups with differential model performance, where splitting decisions are made to maximize heterogeneity in performance between subgroups. We extend these methods to tree ensembles, including both random forests and gradient boosting. Lastly, we illustrate how these ensembles can be used for model combination. We evaluate the methods through simulations and apply them to lung cancer screening data.

\vspace*{0.3in}

\section{Introduction}

The performance of a predictive model is typically assessed using overall measures for the population under consideration. However, model performance often varies across different covariates, resulting in certain subgroups that show superior or inferior performance compared to the overall population \cite{Oakden-Rayner2020HiddenImaging.}. Consequently, overall performance measures may not accurately represent performance within specific subgroups. When model-derived predictions are used for individualized decision making, it is essential that the measures of model performance reflect the subgroup of interest. Traditionally, variation in performance across subgroups is explored by estimating model performance within a limited set of predefined subgroups. The issue of differential model performance is frequently highlighted as a key aspect of fairness, or the lack thereof, in machine learning and artificial intelligence \cite{ricci2022addressing, chen2023algorithmic}.

Data-driven tree-based methods have previously been developed to discover subgroups with differential treatment effects without requiring pre-specification of subgroups \cite{Su2008InteractionData,Foster2011SubgroupData.,Steingrimsson2019SubgroupTrees}. In contrast, methods to identify subgroups with differential model performance are less explored. To the best of our knowledge, the only existing data-driven method for identifying such subgroups is presented in \cite{Foster2017IdentifyingGrowth}. These authors focus specifically on longitudinal data and use splitting decisions based on interaction tests within a shared random effects model. Consequently, their approach applies only to specific data structures, and the validity of their splitting criteria relies on the correct specification of this shared random effects model.

In the first part of this manuscript, we develop two general tree-based methods for the data-driven discovery of subgroups with differential model performance, applicable to any measure of model performance. These methods produce interpretable subgroup structures by recursively partitioning the covariate space into disjoint subgroups that maximize between-subgroup heterogeneity or within-subgroup homogeneity of model performance. Our proposed methods accommodate scenarios where the number of covariates exceeds the sample size and do not require pre-specification of potential subgroups. While the resulting binary tree structure is easily interpretable, it may exhibit high variability. Ensemble methods often outperform single trees in related contexts; thus, we additionally extend our approach to random forest and gradient boosting frameworks to estimate conditional measures of model performance for any given predictive model.

In the second part of the manuscript, we develop ensemble-based methods for model combination, building upon the methods introduced in the first part. There exists extensive literature on model combination, often emphasizing dynamic model selection. For instance, \cite{deSouto2008EmpiricalEnsembles,Brun2016ContributionSelection} estimated model classification error rates in local regions around specific covariate patterns (regions of competence) and selected the optimal model(s) for prediction at these covariate patterns. \cite{Cruz2018DynamicPerspectives} provided a comprehensive overview of dynamic model selection approaches. Typically, regions of competence are defined using algorithms such as clustering or K-Nearest Neighbors. Other methods focused on weighting models include the super learner \cite{vanderLaan2007SuperLearner.} and the mixture of experts \cite{Jacobs1991AdaptiveExperts}. Both methods estimate weights from data (instead of applying fixed rules such as majority voting), but neither is explicitly designed to combine pre-trained predictive models. Moreover, while the super learner provides a uniform set of weights across all covariate patterns, the mixture of experts approach allows for covariate-dependent weights. Our proposed approach can be utilized with any measure of model performance and eliminates the subjective choice regarding the size of the region of competence.



The paper is structured as follows. In Sections~\ref{sec:project2_cart-to} and \ref{sec:project2_PASD}, we propose two extensions of the classification and regression tree algorithm \cite{Breiman1984ClassificationTrees} to identify subgroups with differential model performance.

In Section~\ref{sec:project3_mc_methods}, we extend the single-tree methods to ensemble methods. In Sections~\ref{sec:project3_mc_method1} and \ref{sec:project3_mc_method2}, we apply these ensemble methods to estimate covariate-dependent weights for model combination. Finally, we evaluate the finite-sample performance of the proposed algorithms through simulations and an application to lung cancer screening data.

\section{Notation}

Let $X=(X^{(1)},\dots,X^{(p)})\in\mathcal{X}$ be a $p$-dimensional covariate vector, and $Y\in\mathcal{Y}$ an outcome. A set $w$ is a subgroup if $w\subseteq\mathcal{X}$.  We assume that the data consist of $n$ i.i.d.~observations  $\mathcal{D}=\{(X_i,Y_i):i=1,\dots,n\}$ sampled from the joint distribution of $(X,Y)$. 
We are interested in data-driven discovery of subgroups with differential model performance for a specific measure of model performance and for an already fit prediction model that we denote by $h(X)$. This incorporates prediction models $h(X)$ that were fit using external data and prediction models that were fit using a training set independent of $\mathcal{D}$. 




For reasons that will become clear later, we separate measures of model performance into those that can be calculated for a single observation $i$ and those that cannot. Examples of models performance measures that can be calculated for a single observation $i$ are loss-based measures as for a given loss function $L(Y, h(X))$ the loss associated with observation $i$ is $L(Y_i, h(X_i))$. On the other hand, the area under the ROC curve (AUC) for a subgroup $w$ is defined as 
\[
AUC(w)=\mathbb{E}[I\{h(X)>h(X^\prime)\}\mid X\in w, X^\prime\in w, Y=1,Y^\prime=0],
\]
where $(Y,X)$ and $(Y',X')$ is a random pair of observations from the subgroup $w$ with the former being a case $(Y=1)$ and the latter being a control $(Y'=0)$. As $AUC(w)$ is defined using a pair of observations, it is an example of a measure of model performance that cannot be calculated for each individual observation.

For a given measure of model performance, We use $\mu_i$ to denote the observed model performance for observation $i$ (if available) and $\widehat \mu(w)$ to denote the estimated model performance in subgroup $w$.  For example, for loss-based measures, we have $\mu_i=L(Y_i,h(X_i))$ and $\widehat \mu(w)= \frac{1}{\sum_{i=1}^n I(X_i \in w)} \sum_{i=1}^n I(X_i \in w) L(Y_i,h(X_i))$.

\section{Tree-based methods for discovering subgroups}

In this section we present two methods for discovering subgroups with differential model performance. 

\subsection{Regression tree with transformed outcome}\label{sec:project2_cart-to}

The original Classification and Regression Tree (CART) algorithm \cite{Breiman1984ClassificationTrees} has three main steps: initial tree building, tree pruning, and final tree selection.  The final output of the CART algorithm is a set of subgroups and subgroup-specific estimators where the subgroups are chosen to minimize within-subgroup heterogeneity in the outcome. Following the same principle, we modify the CART algorithm to discover subgroups that minimize within-subgroup heterogeneity in model performance. 

The first step is to grow a large initial tree. The tree-growing process starts with all observations in $\mathcal{D}$ in a single group. For a covariate $X^{(j)}$ and a cut point $c$, define the pair of subgroups $L_{j,c}=\{X\in\mathcal{X}: X^{(j)}\le c\}$ and $R_{j,c}=\{X\in\mathcal{X}: X^{(j)}>c\}$. To make splitting decisions, the algorithm cycles through all possible covariate and split-point pairs $(j,c)$ and finds the pair that optimizes a splitting statistic. Once the optimal pair of covariate and split point combination $(j^*,c^*)$ is found, we split $\mathcal{X}$ into the subgroups $L_{j^*,c^*}$ and $R_{j^*,c^*}$. We recursively apply the above procedure within each resulting subgroup until certain stopping criteria are met, and refer to the resulting tree as the initial tree and denote it by $T_0$.

For discovering subgroups with differential model performance, we propose to use splitting decisions that minimize within subgroup heterogeneity in model performance using some measure of heterogeneity $\sum_{i=1}^n I(X_i \in w) Q(\mu_i, \widehat{\mu}(w))$ (e.g.,~$Q(\mu_i, \widehat{\mu}(w))$ could be the $L_2$ or $L_1$ loss). Hence, the covariate and split-point pair $(j^*,c^*)$ used to split the covariate space into subgroups is selected such that
\begin{equation}
    (j^*,c^*)=\argmin_{j,c}\left[\sum_{X_i\in L_{j,c}}{Q(\mu_i,\widehat{\mu}(L_{j,c}))} + \sum_{X_i\in R_{j,c}}{Q(\mu_i,\widehat{\mu}(R_{j,c}))}\right].\label{eq:cart-to-splitting-criteria}
\end{equation}
Pruning and final tree selection are done in the same way as in the original CART algorithm. That is, we use cost-complexity pruning to first derive a sequence of candidate subtrees and then use cross-validation to select the final tree. Details of pruning, final tree selection and description of the complete algorithm are given in Appendix~\ref{sec:project2_appendix_cart-to-details}. 

One advantage of this algorithm is that it can be implemented by fitting the original CART algorithm on the transformed dataset $\{(X_i,\mu_i): i = 1, \ldots,n\}$. This allows for simple implementation using standard software that implements the CART algorithm. However, a drawback is that it only works with measures of model performance that are available at the individual level. In the next subsection, we propose the Prediction Accuracy Subgroup Discovery (PASD) algorithm a more general tree-based approach that can be used with all univariate measures of model performance.   

\subsection{Prediction Accuracy Subgroup Discovery algorithm}\label{sec:project2_PASD}

The PASD algorithm has three steps, initial tree building, pruning, and final tree selection. 

\subsubsection{Initial tree building} \label{sec:project2_grow_full_tree}

To build a large initial tree, we follow the recursive binary splitting procedure described in Section~\ref{sec:project2_cart-to} 
by maximizing the splitting criterion
\begin{equation}
s_{j,c} = \left(\frac{\widehat{\mu}(L_{j,c}) - \widehat{\mu}(R_{j,c})}{\sqrt{\widehat{\text{Var}}\left[\widehat{\mu}(L_{j,c})-\widehat{\mu}(R_{j,c})\right]}}\right)^2. \label{eq:PASD_split_criterion}
\end{equation}
The splitting criterion $s_{j,c}$ measures the standardized difference in model performance between the two subgroups $L_{j,c}$ and $R_{j,c}$ and using this splitting criterion results in a tree where subgroups are constructed by maximizing the standardized difference between subgroup measures of model performance. When model performance is identical in the subgroups considered, the splitting criterion \eqref{eq:PASD_split_criterion} is asymptotically $\chi^2_1$ distributed. In Appendix~\ref{sec:project2_appendix_acc_estimation}, we demonstrate how the splitting statistic $s_{j,c}$ is calculated for several measures of model performance.


\subsubsection{Pruning the initial tree.} \label{pruning}

The second step is to prune the large, usually overfit, initial tree $T_0$. We propose to create a sequence of candidate trees by recursively removing branches from the initial tree \cite{Leblanc1993SurvivalSplit}. For a tree $T$, let $I_T$ denote the set of internal nodes. A branch $T^{(m)}$ of a tree $T$ is defined as a tree with root node $m\in I_T$, and $T^{(m)}$ contains node $m$ and all descendants of node $m$. 

For a given $\alpha\ge0$, define the split-complexity of a tree $T$ as
\begin{equation}
S_\alpha(T)  \triangleq \sum_{m\in I_T}{s(m)} - \alpha|I_T|, \label{split_complexity}
\end{equation}
where $s(m)$ is the value of the splitting statistic for internal node $m$. The first term in $S_\alpha(T)$ is the sum of the splitting statistics of all internal nodes in the tree $T$. Split complexity $S_\alpha(T)$ measures the trade-off between the subgroup difference in model performance and the complexity of the tree as measured by the number of splits. For a fixed $\alpha$, we say that a subtree of $T_0$ is optimal w.r.t.~split-complexity if it maximizes $S_\alpha(T)$. For a fixed $\alpha$, an additional split in the tree increases the split complexity if the splitting statistic associated with that split is greater than $\alpha$.

To create a sequence of candidate trees that are optimal in terms of split complexity for some $\alpha$ we use the following procedure \cite{Breiman1984ClassificationTrees}. Define the function $g(m;T)$ as
\begin{equation}
g(m;T) \triangleq 
\begin{cases}
\sum_{l\in I_{T^{(m)}}}{s(l)} / |I_{T^{(m)}}|, \quad & m\in I_T \\
+\infty, \quad &\text{otherwise.}
\end{cases}
\label{eq:g_func}
\end{equation}
In the pruning process we first find the internal node $m^*_0=\argmin_{m\in I_{T_0}}{g(m;T_0)}$ and $\alpha_1=g(m^*_0;T_0)$. Node $m^*_0$ is the internal node that is associated with the smallest $\alpha$ value ($\alpha_1$) such that removing the branch becomes preferable compared to keeping it in terms of maximizing split complexity. Let $T_1$ be the resulting tree when branch $T^{(m^*_0)}$ is removed. We then find the internal node $m^*_1=\argmin_{m\in I_{T_1}}{g(m;T_1)}$ and $\alpha_2=g(m^*_1;T_1)$, and obtain $T_2$ by removing branch $T^{(m^*_1)}$ from $T_1$. We repeat this procedure until the remaining tree consists only of the root node. This process results in a sequence of $\alpha$ values $0=\alpha_0<\alpha_1<\ldots<\alpha_{\widehat K}$, with corresponding sequence of nested subtrees $T_0\supset T_1\supset T_2\supset \cdots \supset T_{\widehat K}$ that are candidates for being selected as the final tree.

\subsubsection{Selecting Final Subtree}

We propose two ways of selecting a final tree from the sequence of candidate trees $T_0, T_1, \ldots, T_{\widehat K}$.\\
\noindent \textit{Final tree selection method 1.} For measures of model performance that can be calculated at the individual level, we can perform cross-validation where the tuning parameter $\alpha$ is chosen to minimize the $V$-fold cross-validated prediction error in estimating the measure of model performance.\\
\noindent \textit{Final tree selection method 2.} The second cross-validation method that we propose for final tree selection can also be used with measures of model performance that cannot be calculated at the individual level. In short, the method maximizes the cross-validated split complexity for a fixed $\alpha'$. 


We start by splitting $\mathcal{D}$ into $V$ mutually disjoint and exhaustive sets $\mathcal{D}_{1},\dots,\mathcal{D}_{V}$. For each pair of $v \in \{1, \ldots, V\}$ and $k \in \{0, \ldots, \widehat K\}$, we build a fully grown tree $T_{v,0}$ using the data in $\mathcal{D}\setminus	 \mathcal{D}_v$ and find the optimal subtree of $T_{v,0}$ that corresponds to the penalization parameter $\alpha_k$. For each $v$, this results in $\widehat K+1$ subtrees $T_{v,0}\supseteq T_{v,1}\supseteq T_{v,2}\supseteq \cdots \supseteq T_{v, \widehat K}$. For each subtree, calculate the split-complexity of the tree $S_{\alpha^\prime}(T_{v,k})$ using only observations in $\mathcal{D}_{v}$. So for each pair of $v \in \{1, \ldots, V\}$ and $k \in \{0, \ldots, K\}$, we have a corresponding $S_{\alpha^\prime}(T_{v,k})$. The final  subtree $T_{k^*}$ is then selected as $k^*=\argmax_{k\in\{0,1,\dots,\widehat K\}}{\frac{1}{V}\sum_{v=1}^{V}{S_{\alpha^\prime}(T_{v,k})}}$.
The PASD algorithm with both tree selection methods are presented in Appendix~\ref{sec:project2_appendix_PASD1-details}.

\subsection{Properties of the tree-algorithms}\label{sec:PASD_property}

We introduced three tree-based algorithms above, namely CART with transformed outcome (CART-TO), PASD with final tree selection method one (PASD-1) and PASD with final tree selection method two (PASD-2). Collectively we refer to the tree algorithms for discovering subgroups with differential model performance as PASD. CART-TO and PASD-1 work only with measures of model performance that can be calculated at the individual level, while PASD-2 can be used for any univariate measure of model performance. For all three algorithms, the final tree is a set of terminal nodes and terminal node-specific estimators that are the model performance estimated using data in each terminal node. The following theorem shows that the tree-based algorithms are all $L_2$-consistent. Proof of the theorem is given in Appendix. 

\begin{theorem}
If $Y$ and $h(X)$ are bounded and the number of terminal nodes is of order $o(n/\log n)$ when $n \longrightarrow \infty$, then all three tree algorithms are $L_2$-consistent estimators for conditional model performance at $X$.
\end{theorem}

When we use the same data for tree building and terminal node model performance estimation, the terminal node estimators are likely biased. In order to achieve unbiased terminal node estimators, we can split the data into two independent groups using one part of the data for tree building (subgroup discovery) and the other for terminal node estimation \cite{yang2022causal}. This guarantees that the final subgroup estimators are independent of the subgroup dicovery step. That is, if we let $\widehat{\mu}(\widehat{w})=\frac{\sum_{i=1}^{n_2}{I(X_i\in\widehat{w})\cdot[Y_i-h(X_i)]^2}}{\sum_{i=1}^{m}{I(X_i\in\widehat{w})}}$ denote the accuracy estimator from data $D_2=\{(X_i,Y_i):i=1,\dots,n_2\}$ for final subgroup $\widehat{w}$ obtained using data $D_1=\{(X_i,Y_i):i=1,\dots,n_1\}$, then it is easy to see that conditioned on $D_1$, $\sqrt{n_2}[\widehat{\mu}(\widehat{w})-\mu(\widehat{w})]\to N(0,\sigma^2(\widehat{w}))$. Here we are using squared error as an example but the statement can be easily generalized to other model performance measures.

\section{Ensemble trees for estimation of model performance}\label{sec:project3_PASD_ensemble}


The tree-based algorithms proposed in the last section can be viewed as a prediction model for measures of model performance conditional on $X$ (denoted by $\mu(x)$). To improve performance over single trees, ensemble methods such as random forest \cite{Breiman2001RandomForests} and gradient boosting trees \cite{Friedman2001GreedyMachine} are often used. In this section, we describe extensions to random forests and gradient-boosting trees. Since ensembles of trees no longer have a tree structure, they do not output subgroups of differential performance, and instead, they estimate $\mu(x)$.

The original random forest algorithm \cite{Breiman2001RandomForests} uses fully grown classification and regression trees as building blocks where each individual tree in the forest is fit on a bootstrap sample and only a random subset of covariates is considered when finding the optimal split for each splitting decision. We employ a similar strategy by replacing classification and regression trees with full-grown PASD trees. To avoid overfitting we estimate subgroup-specific model performance of each tree using the out-of-bag (OOB) sample (i.e.,~observations that are not selected in the bootstrap sample). The final prediction is given by the average prediction averaged across all trees. The complete algorithm is given in Supplementary Web Appendix \ref{app:Sec-ensemble}. 



In gradient boosting \cite{Friedman2001GreedyMachine}, multiple trees are fit sequentially with each new tree focusing on where the previous trees did poorly. More specifically, the $m$th tree in the sequence is chosen as $\widehat{T}_m = \argmin_{T}{\sum_{i=1}^{n}{L\left(\mu(x_i),\sum_{u=1}^{m-1}{\widehat{T}_u(x_i)}+T(x_i)\right)}}$
for some user-defined loss function $L$. If $L$ is the squared-error loss, the optimal new tree $\widehat{T}_m$ to be added can be simply found by fitting $\widehat{T}_m$ to the current residuals $\{\mu(x_i)-\sum_{u=1}^{m-1}{\widehat{T}_u(x_i)}:i=1,\dots,n\}$, and the subgroup-specific model performance estimate is the mean $\mu(x_i)$ in that group. For more general loss $L$, the optimal $m$th tree is found by fitting it to the negative gradient of $L$ with respect to the current predicted values evaluated at all the training data points. For example, if $L$ is the absolute error loss, then we fit the $m$th tree to the sign of the current residuals. This approach  is inspired by the gradient descent algorithm in numerical optimization where the minimum of a function is found by going down the loss function space in the direction of the negative gradient.

The total number of trees, $M$, in the model should be carefully chosen as large $M$ can result in overfitting. In addition, each tree is multiplied by a regularization term $\lambda\in(0,1)$ before being added to the sequence. Smaller values of $\lambda$ introduce more regularization, which means that its value should be tuned together with $M$ to prevent overfitting. We present the complete gradient boosting trees algorithm for estimation of model performance in Algorithm~\ref{alg:gb_PASD} in Supplementary Web Appendix \ref{app:Sec-ensemble}.

\section{Model combination using PASD ensembles}\label{sec:project3_mc_methods}

An ensemble of PASD trees estimates the model performance for a given prediction model at a specific covariate pattern. We now describe how an ensemble of PASD trees can be used to combine $K$ prediction models $\{h_1(X),\dots,h_K(X)\}$ by leveraging the strength and weaknesses of each model in different parts of the covariate space.

A well known approach to combining models is Bayesian model averaging which is a linear combination of models with weights proportional to their posterior probability. Our proposed method for combining PASD trees is similar, with the important distinction that the weights are not obtained from a posterior probability distribution. We assume that one of the $K$ models is optimal with respect to some user-defined criteria and we express the uncertainty about our knowledge of which model is optimal using the weights, by extending the formulation in \cite{Raftery2005UsingEnsembles} to incorporate covariate-dependent weights $\{\pi_{k\mid D}(x):k=1,\dots,K\}$ estimated using a separate dataset $D$. The weights $\pi_{k\mid D}(x)$ reflect how well each model $h_k$ performs at $x$ given data $D$ and are constrained to satisfy $\sum_{k=1}^{K}{\pi_{k\mid D}(x)}=1$ and $\pi_{k\mid D}(x)\ge0$ for all $k \in \{1, \ldots, K\}$. In the following two subsections, we propose two new methods for estimating the weights $\pi_{k\mid D}(x)$ using the PASD ensembles.



\subsection{Majority voting model combination}\label{sec:project3_mc_method1}

For a given measure of model performance and a prediction model $h_k(X)$, let $\mu_k(x)$ denote model performance at $x$. Without loss of generality, we assume better model performance is indicated by lower values of $\mu(x)$. Intuitively, we would choose the model in $\{h_1,\dots,h_K\}$ that has the best performance at $x$ to predict the outcome of $x$. That is, we would want to associate each model $h_k$ with weight $I\{\mu_k(x)=\min_{t \in \{1, \ldots, K}{\mu_t(x)}\}$. Given observed data $D$, we propose to estimate the weights $\pi_{k\mid D}(x)$ by fitting multiple PASD trees for each prediction model $h_k(x)$ and see how frequently each model is estimated to have the best predictive performance for a given $x$.


More formally, for each prediction model $h_k$ we fit a random PASD forest using data $D$. Each forest consists of $B$ individual PASD trees $\{\widehat{\mu}_{k}^{(1)},\dots,\widehat{\mu}_{k}^{(B)}\}$, and the weight $\pi_{k\mid D}(x)$ is estimated by
\begin{equation}
    \widehat{\pi}_{k\mid D}(x)=\frac{1}{B}\sum_{b=1}^{B}{I\left\{\widehat{\mu}_{k}^{(b)}(x) = \min_{t \in \{1, \ldots, K\}}{\widehat{\mu}_{t}^{(b)}(x)}\right\}}, \quad k=1,\dots, K \label{eq:mv_weights}
\end{equation}
where $I(A)$ is the indicator function with $I(A)=1$ if $A$ holds and zero otherwise. Hence, we estimate the weight of model $h_k$ at $x$ using the proportion of bootstrap samples where $h_k$ is estimated to have the best performance at $x$ among all models. The prediction from the combined model is given by $\sum_{k=1}^{K}{\widehat{\pi}_{k\mid D}(X)\cdot h_k(X)}$. The complete algorithm is given in Algorithm~\ref{alg:project3_mv_combination} in Supplementary Web Appendix \ref{app:Model-comb}.

Equation (\ref{eq:mv_weights}) is defined for covariate pattern specific measures of model performance $\mu(x)$. For group-level measures of model performance, such as the AUC, the formulation in (\ref{eq:mv_weights}) can be easily adapted so that $\widehat{\mu}_{k}^{(b)}(x)$ represents the estimated model performance for the subgroup that contains $x$ in the tree fitted on the $b$th bootstrap sample for prediction model $h_k$.

To allow for a fair comparison across the $K$ prediction models, the random PASD forest models are fitted on the same bootstrap datasets. Furthermore, since only the predictive mean is of interest, we are directly estimating the weights rather than specifying a parametric form for the individual prediction models and estimating the parameters. The advantage of this approach is that we only need data for random PASD forest fitting and do not require additional data for parameter estimation. The drawback is that we only obtain a point estimate (predictive mean) of the combined predictor without any information on its uncertainty.

\subsection{EM-based model combination}\label{sec:project3_mc_method2}
For the EM-based model combination we specify a parametric form for the weights and use maximum likelihood for parameter estimation. This approach can be viewed both as an extension of the method proposed in \cite{Raftery2005UsingEnsembles} with covariate-dependent weights and as a mixture-of-expert model \cite{Jacobs1991AdaptiveExperts} but with fixed models and weights estimated from the predictive performance of each model. 

We start by using a random PASD forest or a gradient boosting PASD trees model to estimate $\widehat{\mu}_k(x)$. With a continuous outcome, we assume that the density of $Y$ can be written as
$p(y_i\mid x_i;\nu,\gamma) = \sum_{k=1}^{K}{\pi_k(x_i;\nu)\cdot \phi(y_i;h_k(x_i),\sigma_k^2)}$
where $\pi_k(x_i;\nu)=\frac{\exp\{\beta_k\cdot\widehat{\mu}_k(x_i)\}}{\sum_{j=1}^{K}{\exp\{\beta_j\cdot\widehat{\mu}_j(x_i)\}}}$,
and $\phi$ denotes the Gaussian density. The parameters of this model are $\gamma=\{\sigma_1^2,\dots,\sigma_K^2\}$ and  $\nu=\{\beta_1,\dots,\beta_K\}$. We use a latent variable representation by defining latent variables $Z_i$ such that $Y\mid X,Z=k\sim N(h_k(X),\sigma_k^2)$, and $Pr[Z=k\mid X]=\pi_k(X;\nu)$. We propose to use the EM algorithm \cite{Dempster1977MaximumAlgorithm} to iteratively find the maximum likelihood estimates of the parameters. The details are given in Algorithm~\ref{alg:project3_numerical_em} and Supplementary Web Appendix~\ref{sec:project3_appendix_em_details}.

\begin{algorithm}
\caption{Model combination with EM}\label{alg:project3_numerical_em}
\nl Initialize parameter estimates $\gamma^{(0)}$\;
\For{t=1,2,\dots}{
\begin{enumerate}
    \item E-step: calculate 
    \begin{equation*}
        \lambda_{ik}=\frac{\phi\left(y_i;h_k(x_i),{\sigma_k^2}^{(t-1)}\right)\cdot \pi_k(x_i;\nu^{(t-1)})}{\sum_{j=1}^{K}{\phi\left(y_i;h_j(x_i),{\sigma_j^2}^{(t-1)}\right)\cdot \pi_j(x_i;\nu^{(t-1)})}}, \quad i=1,\dots,n, \quad k=1,\dots,K
    \end{equation*}
    \item M-step: 
    \begin{align*}
        {\sigma_j^2}^{(t)} 
        &= \frac{\sum_{i=1}^{n}{\lambda_{ij}[h_j(x_i)-y_i]^2}}{\sum_{i=1}^{n}{\lambda_{ij}}}, \quad j=1,\dots,K \\
        \nu^{(t)} &= \argmax_{\nu}{\sum_{i=1}^{n}{\sum_{k=1}^{K}{\lambda_{ik}\log \pi_k(x_i;\nu)}}}.
    \end{align*}
\end{enumerate}
}
\end{algorithm}

There is no analytic solution in the M-step of Algorithm~\ref{alg:project3_numerical_em} for $\nu^{(t)}$, and numerical optimization techniques are used to perform maximization in this step. 
Since the $\nu$ parameter updates are obtained by maximizing a term that resembles the log-likelihood of multinomial generalized linear models (by treating $\lambda_{ik}$ as the outcome in the M-step of Algorithm~\ref{alg:project3_numerical_em}), we derive in Appendix~\ref{sec:project3_appendix_fisherScoring} the Fisher scoring update rules for updating $\nu$.  

Another solution to avoiding the inner loop of maximization was proposed by \cite{Xu1994AnExperts} where the weights are specified in a different parametric form. In our case, this suggests
\begin{equation}
    \pi_k(x_i;\nu) = \frac{\beta_k \cdot p(\widehat{\mu}_k(x_i);\tau^2)}{p(x_i;\nu)}=\frac{\beta_k \cdot p(\widehat{\mu}_k(x_i);\tau^2)}{\sum_{j=1}^{K}{\beta_j\cdot p(\widehat{\mu}_j(x_i);\tau^2)}}
\end{equation}
with $\sum_{k=1}^{K}{\beta_k}=1$ and $\beta_k\ge0$. If $\widehat{\mu}_k(x)$ is estimating squared error, we further assume $p(\widehat{\mu}_k(x);\tau^2)$ is the density of a scaled chi-squared $\tau^2\chi^2_1$ given by
$p(\widehat{\mu}_k(x);\tau^2)=\frac{1}{\tau^2\sqrt{2\pi}}\left(\frac{\widehat{\mu}_k(x)}{\tau^2}\right)^{-1/2}\exp\left\{-\frac{\widehat{\mu}_k(x)}{2\tau^2}\right\}$. This allows analytic expression for parameter updates in the M-step. However, this formulation imposes assumptions on the distribution of $p(x_i;\nu)$. 
The details of this algorithm with analytic solutions for the M-step parameter updates are given in Algorithm~\ref{alg:project3_analytic_em} in Appendix~\ref{sec:project3_appendix_em_details}.

\section{Simulations}\label{sec:project2_simulation}

In this section, we used simulations to a) evaluated the performance of the three algorithms for subgroup identification, b) compare the estimation accuracy of a single PASD tree to PASD ensembles, and c) demonstrate how PASD ensembles can be used to combine a set of prediction models.

\subsection{Evaluating the performance of three subgroup identification tree algorithms}

\paragraph{Simulation setup}

The covariate vector was six dimensional where the first four components were simulated from a multivariate normal distribution with mean zero and covariance matrix with diagonal elements equal to one and off-diagonal elements equal to $c$, where $c$ varied across the simulation settings. The last two components of the covariate vector were simulated from a Bernoulli distribution with parameters $0.5$ and $0.7$. 

We considered four different simulation settings. The outcome $Y$ was continuous and was simulated as $Y=2+X^{(1)}-(X^{(2)})^2+I(X^{(3)}>0)+1.5X^{(5)}+1.5X^{(2)}X^{(5)}+\epsilon$. Table~\ref{tab:project2_sim_desc} provides details on how the risk prediction model $h(X)$, the error distribution $\epsilon$, and the off-diagonal elements of the covariance matrix $c$ were selected. We use mean squared error (MSE) as the measure of model performance.

\begin{table}[!htb]
\centering
\caption{Prediction model, error and covariance specifications of the four simulation settings.}
\label{tab:project2_sim_desc}
\resizebox{\textwidth}{!}{%
\begin{tabular}{@{}clc@{}}
\toprule
Setting & \multicolumn{1}{c}{prediction model $h(X)$} & Error ($\varepsilon$) and Covariance (c) \\ \midrule
1 & $h=2+X^{(1)}-(X^{(2)})^2+I(X^{(3)}>0)+1.5X^{(5)}+1.5X^{(2)}X^{(5)}$ & $\epsilon\sim N(0,4), c=0$ \\
2 & $h$ $=2+X^{(1)}-(X^{(2)})^2+0.5X^{(5)}+1.5X^{(2)}X^{(5)}$ & $\epsilon\sim N(0,4), c=0$ \\
3 & $h$ $=2+X^{(1)}-(X^{(2)})^2+0.5X^{(5)}+1.5X^{(2)}X^{(5)}$ & $\epsilon\sim N(0,4), c=0.3$ \\
4 & $h=2+X^{(1)}-(X^{(2)})^2+I(X^{(3)}>0)+1.5X^{(5)}+1.5X^{(2)}X^{(5)}$ & $\epsilon\sim N(0,(X^{(6)}/2+1)^2), c=0$ \\ \bottomrule
\end{tabular}%
}
\end{table}

Differential model performance occurs when some observations are inherently harder to predict than others, and it can also be a consequence of a misspecified prediction model where the impact of misspecification on model performance is larger in some subgroups than others.

\begin{itemize}
    \item In simulation setting 1, the error distribution $\epsilon$ does not depend on covariates and the prediction model $h$ is correctly specified. Thus, the MSE does not vary as a function of $X$ and the true tree makes no splits (see Figure~\ref{img:sim_case1_true_tree} in Appendix~\ref{sec:project2_appendix_true_tree_structure} for the true tree). 
    
    \item In simulation setting 2, $h$ is incorrectly specified when $X^{(3)}>0$ and/or $X^{(5)}=1$. This mispecification results in worse model performance in parts of the covariate space where $X^{(3)}>0$ and/or $X^{(5)}=1$. Figure~\ref{img:sim_case2_true_tree} in Appendix~\ref{sec:project2_appendix_true_tree_structure} shows the true tree that partitions the covariate space into four different subgroups depending if $X^{(3)}>0$ and/or $X^{(5)}=1$.  
    
    \item Simulation setting $3$ is identical to simulation setting $2$ except that we set the off-diagonal element of the covariance matrix for the continuous component of the covariate vector to $c=0.3$. The true subgroups in this simulation setting were the same as in simulation setting~$2$.
    \item In simulation setting 4, the error term, which is equal to the MSE of the true model, in the true data generating process depends on the binary covariate $X^{(6)}$. Thus for a correctly specified model, observations with $X^{(6)}=1$ have higher MSE than observations with $X^{(6)}=0$. Hence, the correct tree splits on whether $X^{(6)}=1$ or not (Figure~\ref{img:sim_case4_true_tree} in Appendix~\ref{sec:project2_appendix_true_tree_structure} shows the correct tree). 
\end{itemize}

For all four settings we ran 1000 simulations and for each simulation we sampled 1000 observations from the joint distribution of $(X^{(1)},\dots,X^{(6)},Y)$. For all settings, the prediction model $h(X)$ is held fixed using the specification listed in Table \ref{tab:project2_sim_desc}. To evaluate the performance of the different methods for subgroup discovery we compare them in terms of mean squared error (MSE), pairwise prediction similarity (PPS), proportion of fitted trees that only split on the covariates that affect prediction performance, and proportion of fitted trees that only split on the correct covariates and have an equal number of subgroups compared to the true tree. In section \ref{app:eval-sim} in Supplementary Web Appendix \ref{app:sim} we provide more details on the evaluation measures.

\paragraph{Results}
Table~\ref{tab:project2_sim_results_alphaPrime_4_simpleVersion} presents the simulation results. For simulation setting 1, the PASD-2 algorithm correctly generates a tree with only the root node 97.8\% of the time. Compared with CART-TO and PASD-1, it has the smallest percentage of splitting on noise variables. In fact, in all settings, the PASD-2 algorithm is better than other algorithms in terms of the proportion of times it avoided splitting on noise variables. 

However, as we can see, when the prediction model is specified incorrectly (setting 2 and setting 3), the PASD-2 algorithm is more likely to underfit than CART-TO and PASD-1 reflected by the higher MSE and lower PPS values. For all algorithms, underfitting occurs due to the correct covariates not being sufficiently split. This means that even if the algorithms can produce trees that discover the correct covariates, they will sometimes be overly pruned, eventually resulting in fewer splits than needed. This could be a result of how we set the value of $\alpha^\prime$ when we select the final subtree. In our current implementation, $\alpha^\prime$ is defaulted to $4$ which penalizes tree complexity more compared to $\alpha^\prime=2$ or $3$. Additional simulation results with $\alpha^\prime$ being $2,3$ and log of test data size are provided in Supplementary Web Appendix~\ref{sec:project2_appendix_additional_sim_results} where we see increasing $\alpha^\prime$ leads to less overfitting but in some cases more underfitting.

\begin{table}[!thb]
\centering
\caption{Results of the four simulation settings with $\alpha^\prime=4$. Method refers to the subgroup identification methods, MSE is mean squared error, PPS is pairwise prediction similarity.}
\label{tab:project2_sim_results_alphaPrime_4_simpleVersion}
\begin{tabular}{@{}cccccc@{}}
\toprule
Setting & Method & MSE & PPS & No noise & Correct Model. \\ \midrule
1       & CART  & 32.095   & 0.976    & 0.952    & 0.952     \\
1       & PASD1 & 32.110   & 0.961    & 0.915    & 0.915     \\
1       & PASD2 & 32.071   & 0.993    & 0.978    & 0.978     \\ \midrule
2       & CART  & 57.077   & 0.765    & 0.692    & 0.043     \\
2       & PASD1 & 57.134   & 0.780    & 0.521    & 0.074     \\
2       & PASD2 & 57.681   & 0.494    & 0.945    & 0.032     \\ \midrule
3       & CART  & 56.647   & 0.769    & 0.676    & 0.037     \\
3       & PASD1 & 56.731   & 0.772    & 0.502    & 0.072     \\
3       & PASD2 & 57.086   & 0.510    & 0.924    & 0.028     \\ \midrule
4       & CART  & 7.950    & 0.977    & 0.913    & 0.913     \\
4       & PASD1 & 7.953    & 0.968    & 0.864    & 0.864     \\
4       & PASD2 & 7.937    & 0.993    & 0.957    & 0.957     \\ \bottomrule
\end{tabular}
\end{table}


\subsection{Comparing the performance of a single PASD tree and PASD ensembles}\label{sec:project3_sim_PASD_ensemble}

In this subsection we used simulations to evaluate how accurately PASD ensembles estimate $\mu(x)$ and compared it to single PASD trees. To simulate the data we started with a 10-dimensional covariate vector with elements simulated from a $Uni(0,1)$. The outcome was simulated using $Y=10\sin(\pi X_1X_2)+20(X_3-0.5)^2+10X_4+5X_5+\epsilon$ where $\epsilon\sim N(0,2X_4)$. Only five of the covariates are related to the outcome and the remaining ones are independent of $Y$. This example is taken from \cite{Friedman1991MultivariateSplines}, but we modified the error distribution to make it dependent on one of the covariates. 

\begin{figure}[!htbp]
    \centering
    \begin{subfigure}[b]{0.44\textwidth}
         \centering
         \includegraphics[width=\textwidth]{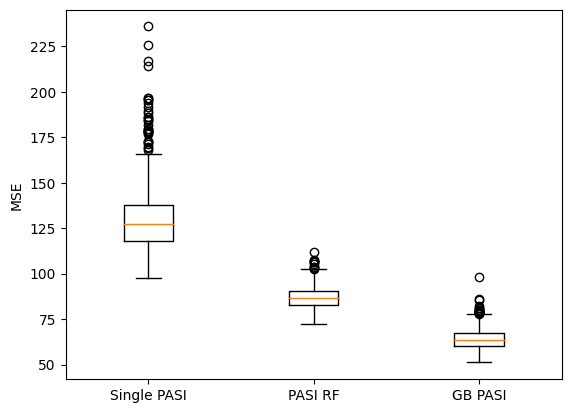}
         \caption{Linear Regression}
         \label{fig:mse_sglvsEnsem_Friedman1_lm}
     \end{subfigure}
     \qquad
     \begin{subfigure}[b]{0.45\textwidth}
         \centering
         \includegraphics[width=\textwidth]{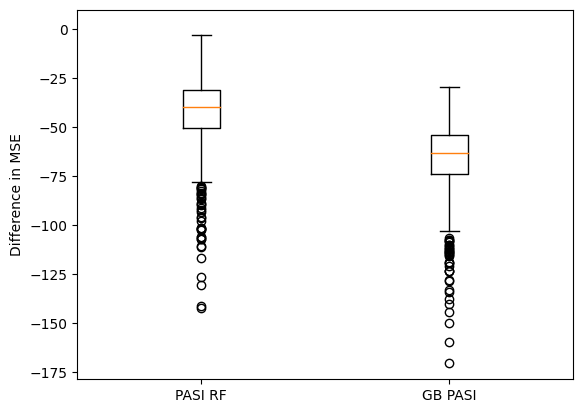}
         \caption{Linear Regression}
         \label{fig:relative_mse_sglvsEnsem_Friedman1_lm}
     \end{subfigure}
     \begin{subfigure}[b]{0.44\textwidth}
         \centering
         \includegraphics[width=\textwidth]{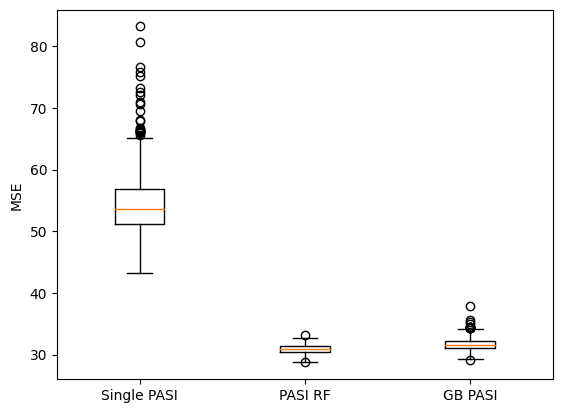}
         \caption{Random Forest}
         \label{fig:mse_sglvsEnsem_Friedman1_rf}
     \end{subfigure}
     \qquad
     \begin{subfigure}[b]{0.45\textwidth}
         \centering
         \includegraphics[width=\textwidth]{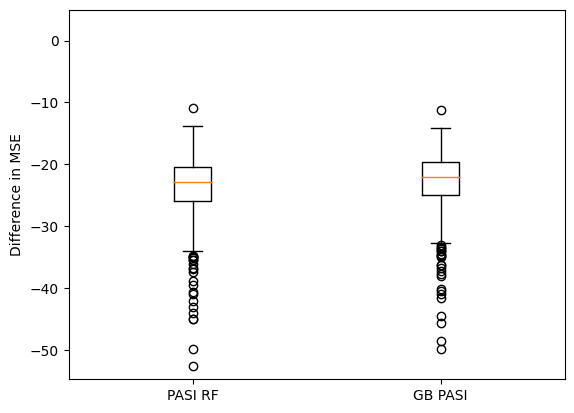}
         \caption{Random Forest}
         \label{fig:relative_mse_sglvsEnsem_Friedman1_rf}
     \end{subfigure}
    \caption{Mean squared error (MSE) of single PASD tree, random PASD forest (RF), and gradient boosting (GB) PASD trees when they are used to estimate the conditional mean squared errors of a linear regression and a random forest model. Figures a) and c) show the results when evaluating a linear regression and a random forest model, respectively. Figures b) and d) show results when for each simulation the mean squared error of the PASD tree is subtracted from the mean squared error of the PASD random forest and the gradient boosting PASD trees. For all simulations, both PASD ensembles achieved better performance (lower MSE) than the single PASD tree.}
    \label{fig:mse_sglvsEnsem_Friedman1}
\end{figure}

We considered two models for the prediction model being evaluated: a linear regression and a random forest. The evaluation measure $\mu(X)$ is the mean squared error. For each model, we fit (a) a single PASD tree, (b) a random PASD forest, and (c) gradient boosting PASD trees on a dataset $D_{tree}$ of size 1000. We compared the performance of the single PASD tree and the PASD ensembles using the MSE of the mean squared error on an independent dataset of size $n_{test} = 10000$ simulated using the same data generating distribution. More precisely, if $\mu(x)$ is the true conditional mean squared error and $\widehat{\mu}(X)$ is the estimated conditional mean squared error than the evaluation is given by $\frac{1}{n_{test}}\sum_{i=1}^{n_{test}}{[\mu(X_i)-\widehat{\mu}(X_i)]^2}$. We ran 1000 simulations and the results are presented in Figure~\ref{fig:mse_sglvsEnsem_Friedman1}.

Figure~\ref{fig:mse_sglvsEnsem_Friedman1_lm} shows the distributions of the MSE of the single PASD tree, random PASD forest and gradient boosting PASD trees in estimating the squared errors of the linear regression model. The single PASD tree had higher MSEs compared to both PASD ensembles, with gradient boosting PASD trees generally having the lowest MSE. Figure~\ref{fig:relative_mse_sglvsEnsem_Friedman1_lm} looks at the MSE difference by subtracting the MSE of the single PASD tree. That is, in each of the 1,000 simulations of the experiment, we calculate the MSE of the PASD ensembles and subtract the single PASD tree MSE from them. It is clear that the PASD ensembles always had lower MSE (i.e., were always better) than the single tree in all of the 1000 simulations as indicated by the negative differences.  Figures~\ref{fig:mse_sglvsEnsem_Friedman1_rf} and \ref{fig:relative_mse_sglvsEnsem_Friedman1_rf} give similar conclusions in the case when random forest is the prediction model being evaluated.

\subsection{Effectiveness of proposed model combination methods}\label{sec:project3_sim_mc}

In this section, we demonstrate the effectiveness of the proposed model combination methods described in Section \ref{sec:project3_mc_methods}.  

\paragraph{Simulation setup}

The covariate vector was six dimensional where the first four components were simulated from a multivariate normal distribution with mean zero and covariance matrix with diagonal elements equal to one and off-diagonal elements equal to $0.3$. The last two components of the covariate vector were simulated from a Bernoulli distribution with parameters $0.5$ and $0.7$. The binary outcome $Y$ was generated from the logistic regression model $Y=I(\eta>0)$ with $\eta=2+X_1-1.5X_2+X_2^2-2I(X_3>0)+1.5X_4-2X_4^3+X_5+0.5X_2X_5+\epsilon$
where $\epsilon\sim\text{Logistic}(0,1)$. 

We considered combining four prediction models for the conditional probability of $Y$ given $X$ that are detailed in Table~\ref{tab:predModels_for_ModelCombSim}. The four logistic regression models were chosen such that they all have similar overall AUC (ranging between 0.68 and 0.69) but each of the models is somewhat misspecified. For example, Model 1 did not include $X_2$ and $X_3$, and Model 2 misspecified the order of the polynomial term of $X_4$ and included the noise variable $X_6$.

\begin{table}[!htb]
\centering
\caption{Logistic regression models that are used in simulation to evaluate the effectiveness of proposed model combination methods.}
\label{tab:predModels_for_ModelCombSim}
\begin{tabular}{@{}c|l@{}}
\toprule
  & \multicolumn{1}{c}{Prediction Model $h(X)=\text{logit}^{-1}(\widehat{\eta})$}                                                                 \\ \midrule
1 & $\widehat{\eta}=2+2X_1+1.5X_4-X_4^3+2X_5+2.5X_2X_5$                                      \\
2 & $\widehat{\eta}=2+X_1-1.5X_2+X_2^2-2I(X_3>0)+1.5X_4-X_4^2+X_5+1.5X_2X_5+0.5X_6$ \\
3 & $\widehat{\eta}=2+X_1-1.5X_2+X_2^2+1.5X_4-1.5X_4^2+X_5+X_2X_5$ \\
4 & $\widehat{\eta}=2+X_1-2I(X_3>0)+1.5X_4-X_4^3+0.8X_5+X_2X_5+2.5X_6$ \\ \bottomrule
\end{tabular}%
\end{table}

\begin{figure}[!htb]
    \centering
    \includegraphics[scale=0.55]{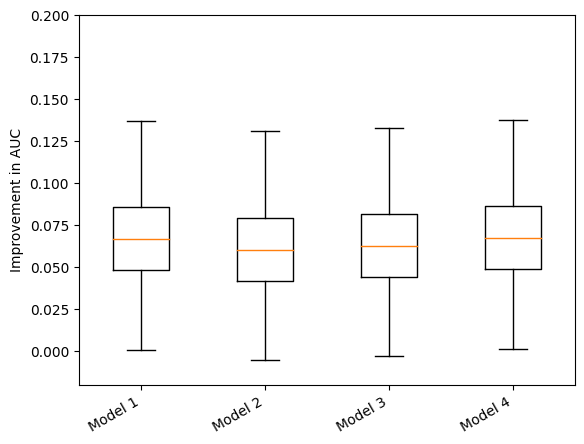}
    \caption{Difference in test set AUC between the combined model and the four individual models. A positive difference indicates that the combined model has a higher AUC compared to the single model.}
    \label{fig:auc_combVSindiv_simulation}
\end{figure}

We simulated a sample $D_{tree}$ of size 1000 from the data generating distribution and applied Algorithm~\ref{alg:project3_mv_combination} to combine the four prediction models with weights estimated using $D_{tree}$. We evaluated and compared the AUC of the combined model with the four individual models using a separate test set $D_{test}$ of size 10,000, and repeated the procedure 1000 times to produce Figure~\ref{fig:auc_combVSindiv_simulation}. In all simulations, the combined model was almost always better than the four individual models. The average AUC of the combined model was 0.74 indicating an improvement over the original prediction models (which ranged from 0.68 to 0.69).

\section{Analysis of lung cancer screening data}

In this section we use data from the National Lung Cancer Screening trial (NLST) to evaluate the methods developed \cite{national2011reduced}. NLST was a large randomized trial that showed a substantial reduction in lung cancer specific mortality when screened with CT screening compared to chest X-ray screening. 

\subsection{Discovery of sbugroups with differential model performance}
\label{sec:project2_nlst_analysis}


The risk prediction model we considered is the PLCOm2012 model, a logistic regression model that predicts the six year risk of developing lung cancer \cite{Tammemagi2013SelectionScreening}, that was developed using data from the Prostate, Lung, Colorectal, and Ovarian (PLCO) Cancer Screening Trial. We applied the PASD tree algorithm to discover subgroups of the NLST data that have differential predictive performance  Covariates that were used as inputs to the algorithm are age, education, body mass index (BMI), chronic obstructive pulmonary disease (COPD), race/ethnicity, personal history of cancer, family history of lung cancer, smoking status, smoking intensity, smoking duration, and smoking quit-time.

The original NLST dataset contains information for 53,452 participants. In order to apply the PLCOm2012 risk prediction model, we preprocessed the raw data to remove ineligible and missing cases and ended up with 48,909 complete observations. We only considered covariates used by the PLCOm2012 model as inputs to the PASD tree algorithm and set the maximum depth of the tree to be three and the minimum number of samples in the terminal nodes to be 500. We used AUC as the model performance measure, and since the cross-validation step in final tree selection randomly splits the data, we considered 1000 different seeds for this process to account for the variability. As tree-based methods are data-adaptive, they can be over-optimistic in searching for the differences in model performance. Thus, to evaluate the tree as a prediction model for risk model performance (in this case, AUC), we split the NLST data into training and test, fit the PASD tree on the training part and evaluated the AUC estimations on the test set. 

The most frequently selected tree (70\% of the time) is the root-node only tree where all observations fall in a single group where the AUC of that group on the test set is $0.68$. This suggests that there is not substantial heterogenity in AUC across subgroups. In Supplementary Web Appendix \ref{sec:project2_compas_analysis} we present analysis of the COMPASS dataset where several subroups with differential model performance were identified.

\subsection{Model combination with PASD ensembles}

In this section, we evaluate the proposed PASD ensemble methods and demonstrate their application in model combination using the NLST data. We considered four previously developed risk models for lung cancer where the outcome was being diagnosed with lung cancer in six years from study enrollment. They were: The Bach model \cite{Bach2003VariationsSmokers.}, the PLCOm2012 model \cite{Tammemagi2013SelectionScreening}, the PLCOm2012 model excluding the race/ethnicity covariate \cite{Pasquinelli2020RiskDisparities.}, and the simplified PLCOm2012 model that only includes age and smoking history as predictors \cite{tenHaaf2017RiskStudy}. The covariates used in each of the four models are given in Table~\ref{tab:project3_four_model_covariates} in Appendix~\ref{sec:project3_appendix_lcModels_additional_info}. All variables are available in the NLST data except asbestos exposure. However, as the number of participants with exposure to asbestos in the NLST data is expected to be very low \cite{tenHaaf2017RiskStudy}, we assumed all individuals in NLST had no asbestos exposure.

In Supplementary Web Appendix \ref{app:nlst-ensemble} we compare single PASD trees and PASD ensembles and now we demonstrate the methods for model combination proposed in Section~\ref{sec:project3_mc_methods} by applying them to the NLST data to estimate weights for combining the prediction models M1-M4. The data were split into a training set and a test set of sizes $9781$ and $39128$, respectively. The sample correlation between the predicted risks given by any pair of models is strong (at least $0.8$ when estimated using the test set; see Table~\ref{tab:project3_corr_M1_M4} in Supplementary Web Appendix~\ref{sec:project3_appendix_lcModels_additional_info}).

\begin{figure}[!htbp]
    \centering
    \begin{subfigure}[b]{0.45\textwidth}
         \centering
         \includegraphics[width=\textwidth]{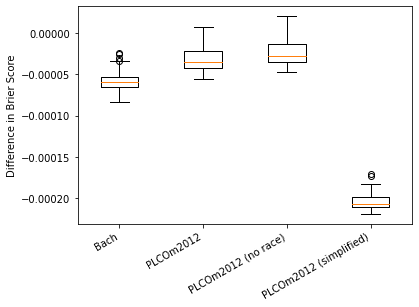}
         \caption{Brier Score}
         \label{fig:brierScore_combVSindiv_nlst}
     \end{subfigure}
     \qquad
     \begin{subfigure}[b]{0.45\textwidth}
         \centering
         \includegraphics[width=\textwidth]{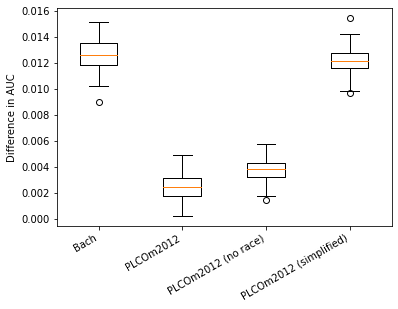}
         \caption{AUC}
         \label{fig:auc_combVSindiv_nlst}
     \end{subfigure}
    \caption{Comparison between combined model and individual risk models in terms of Brier Score and area under the curve (AUC). Each boxplot shows the difference between the performance of combined model and the model listed on the x-axis  across 100 different splits into a training and a test set. A difference in Brier score lower than zero implies better performance of the combined model and a difference in AUC greater than zero implies better performance of the combined model.}
    \label{fig:combVSindiv_nlst}
\end{figure}

We applied the majority voting based model combination method (Algorithm~\ref{alg:project3_mv_combination}) using the training data and evaluated the performance using the test data. The measures of model performance we considered were the Brier score and AUC. We repeated the procedure 100 times and compared the Brier score of the combined model to the four individual prediction models. The results are given in Figure~\ref{fig:brierScore_combVSindiv_nlst}. The Brier score of the combined model was lower than any of the individual prediction models most of the time. The improvement was the largest for the PLCOm2012 simplified model. In this example, since we have a binary outcome ($Y\in\{0,1\}$) and the predicted probabilities from the prediction models were very small, the Brier score was dominated by observations with $Y=1$. That is, $\frac{1}{n}\sum_{i:y_i=1}{(1-\widehat{p}_i)^2}$ contributed most (more than 95\%) of the total Brier score $\frac{1}{n}\sum_{i=1}^{n}{(y_i-\widehat{p}_i)^2}$ where $\widehat{p}_i$ is the predicted probability for participant $i$ from the risk model. The results in Figure~\ref{fig:auc_combVSindiv_nlst} are similar when AUC was used as the performance measure to combine models. We see larger improvements for the Bach and PLCOm2012 simplified models.

\section{Discussion}

By extending the popular classification and regression tree algorithm, we developed three novel tree-based algorithms for discovery of subgroups with differential model performance.  We then extended these algorithms to random forest and gradient boosting methods for estimating model performance for a given covariate pattern and developed methods that use the ensemble method for model combination. 

A useful and a distinct feature of single tree structures is their interpretability. However, they come with their set of challenges, particularly high variability. An important disadvantage of employing an ensemble of PASD trees is the loss of the inherent interpretability observed in single PASD trees. Future research directions should look into developing variable importance measures to understand the relative contribution of each covariate in constructing the trees, potentially offering a means to offset the interpretability loss from ensemble methods. Moreover, as ensemble models continue to gain traction, post-processing methods to extract interpretable subgroups from the ensemble outputs deserve exploration.


\bibliographystyle{ieeetr}
\bibliography{reference}

\begin{thebibliography}{10}

\bibitem{Oakden-Rayner2020HiddenImaging.}
L.~Oakden-Rayner, J.~Dunnmon, G.~Carneiro, and C.~R{\'{e}}, ``{Hidden Stratification Causes Clinically Meaningful Failures in Machine Learning for Medical Imaging.},'' {\em Proceedings of the ACM Conference on Health, Inference, and Learning}, vol.~2020, pp.~151--159, 4 2020.

\bibitem{ricci2022addressing}
M.~A. Ricci~Lara, R.~Echeveste, and E.~Ferrante, ``Addressing fairness in artificial intelligence for medical imaging,'' {\em nature communications}, vol.~13, no.~1, p.~4581, 2022.

\bibitem{chen2023algorithmic}
R.~J. Chen, J.~J. Wang, D.~F. Williamson, T.~Y. Chen, J.~Lipkova, M.~Y. Lu, S.~Sahai, and F.~Mahmood, ``Algorithmic fairness in artificial intelligence for medicine and healthcare,'' {\em Nature biomedical engineering}, vol.~7, no.~6, pp.~719--742, 2023.

\bibitem{Su2008InteractionData}
X.~Su, T.~Zhou, X.~Yan, J.~Fan, and S.~Yang, ``{Interaction trees with censored survival data},'' {\em The international journal of biostatistics}, vol.~4, p.~2, 1 2008.

\bibitem{Foster2011SubgroupData.}
J.~C. Foster, J.~M.~G. Taylor, and S.~J. Ruberg, ``{Subgroup identification from randomized clinical trial data.},'' {\em Statistics in medicine}, vol.~30, pp.~2867--2880, 10 2011.

\bibitem{Steingrimsson2019SubgroupTrees}
J.~A. Steingrimsson and J.~Yang, ``{Subgroup identification using covariate-adjusted interaction trees},'' {\em Statistics in Medicine}, vol.~38, pp.~3974--3984, 9 2019.

\bibitem{Foster2017IdentifyingGrowth}
J.~C. Foster, D.~Liu, P.~S. Albert, and A.~Liu, ``{Identifying subgroups of enhanced predictive accuracy from longitudinal biomarker data using tree-based approaches: applications to fetal growth},'' {\em Journal of the Royal Statistical Society. Series A, (Statistics in Society)}, vol.~180, pp.~247--261, 1 2017.

\bibitem{deSouto2008EmpiricalEnsembles}
M.~C.~P. de~Souto, R.~G.~F. Soares, A.~Santana, and A.~M.~P. Canuto, ``{Empirical comparison of Dynamic Classifier Selection methods based on diversity and accuracy for building ensembles},'' in {\em 2008 IEEE International Joint Conference on Neural Networks (IEEE World Congress on Computational Intelligence)}, pp.~1480--1487, 2008.

\bibitem{Brun2016ContributionSelection}
A.~L. Brun, A.~S. Britto, L.~S. Oliveira, F.~Enembreck, and R.~Sabourin, ``{Contribution of data complexity features on dynamic classifier selection},'' in {\em 2016 International Joint Conference on Neural Networks (IJCNN)}, pp.~4396--4403, 2016.

\bibitem{Cruz2018DynamicPerspectives}
R.~M.~O. Cruz, R.~Sabourin, and G.~D.~C. Cavalcanti, ``{Dynamic classifier selection: Recent advances and perspectives},'' {\em Information Fusion}, vol.~41, pp.~195--216, 2018.

\bibitem{vanderLaan2007SuperLearner.}
M.~J. van~der Laan, E.~C. Polley, and A.~E. Hubbard, ``{Super learner.},'' {\em Statistical applications in genetics and molecular biology}, vol.~6, p.~Article25, 2007.

\bibitem{Jacobs1991AdaptiveExperts}
R.~A. Jacobs, M.~I. Jordan, S.~J. Nowlan, and G.~E. Hinton, ``{Adaptive Mixtures of Local Experts},'' {\em Neural Computation}, vol.~3, no.~1, pp.~79--87, 1991.

\bibitem{Breiman1984ClassificationTrees}
L.~Breiman, J.~H. Friedman, R.~A. Olshen, and C.~J. Stone, {\em {Classification and Regression Trees}}.
\newblock Monterey, CA: Wadsworth and Brooks, 1984.

\bibitem{Leblanc1993SurvivalSplit}
M.~Leblanc and J.~Crowley, ``{Survival Trees by Goodness of Split},'' {\em Journal of the American Statistical Association}, vol.~88, pp.~457--467, 6 1993.

\bibitem{yang2022causal}
J.~Yang, I.~J. Dahabreh, and J.~A. Steingrimsson, ``Causal interaction trees: Finding subgroups with heterogeneous treatment effects in observational data,'' {\em Biometrics}, vol.~78, no.~2, pp.~624--635, 2022.

\bibitem{Breiman2001RandomForests}
L.~Breiman, ``{Random Forests},'' {\em Machine Learning}, vol.~45, no.~1, pp.~5--32, 2001.

\bibitem{Friedman2001GreedyMachine}
J.~Friedman, ``{Greedy Function Approximation : A Gradient Boosting Machine},'' {\em The Annals of Statistics}, vol.~29, no.~5, pp.~1189--1232, 2001.

\bibitem{Raftery2005UsingEnsembles}
A.~E. Raftery, T.~Gneiting, F.~Balabdaoui, and M.~Polakowski, ``{Using Bayesian model averaging to calibrate forecast ensembles},'' {\em Monthly Weather Review}, vol.~133, no.~5, pp.~1155--1174, 2005.

\bibitem{Dempster1977MaximumAlgorithm}
A.~P. Dempster, N.~M. Laird, and D.~B. Rubin, ``{Maximum Likelihood from Incomplete Data Via the EM Algorithm},'' {\em Journal of the Royal Statistical Society: Series B (Methodological)}, vol.~39, no.~1, pp.~1--22, 1977.

\bibitem{Xu1994AnExperts}
L.~Xu, M.~Jordan, and G.~E. Hinton, ``{An Alternative Model for Mixtures of Experts},'' in {\em Advances in Neural Information Processing Systems} (G.~Tesauro, D.~Touretzky, and T.~Leen, eds.), vol.~7, MIT Press, 1994.

\bibitem{Friedman1991MultivariateSplines}
J.~H. Friedman, ``{Multivariate Adaptive Regression Splines},'' {\em The Annals of Statistics}, vol.~19, no.~1, pp.~1--67, 1991.

\bibitem{national2011reduced}
N.~L. S. T.~R. Team, ``Reduced lung-cancer mortality with low-dose computed tomographic screening,'' {\em New England Journal of Medicine}, vol.~365, no.~5, pp.~395--409, 2011.

\bibitem{Tammemagi2013SelectionScreening}
M.~C. Tammem{\"{a}}gi, H.~A. Katki, W.~G. Hocking, T.~R. Church, N.~Caporaso, P.~A. Kvale, A.~K. Chaturvedi, G.~A. Silvestri, T.~L. Riley, J.~Commins, and C.~D. Berg, ``{Selection Criteria for Lung-Cancer Screening},'' {\em New England Journal of Medicine}, vol.~368, pp.~728--736, 2 2013.

\bibitem{Bach2003VariationsSmokers.}
P.~B. Bach, M.~W. Kattan, M.~D. Thornquist, M.~G. Kris, R.~C. Tate, M.~J. Barnett, L.~J. Hsieh, and C.~B. Begg, ``{Variations in lung cancer risk among smokers.},'' {\em Journal of the National Cancer Institute}, vol.~95, pp.~470--478, 3 2003.

\bibitem{Pasquinelli2020RiskDisparities.}
M.~M. Pasquinelli, M.~C. Tammem{\"{a}}gi, K.~L. Kovitz, M.~L. Durham, Z.~Deliu, K.~Rygalski, L.~Liu, M.~Koshy, P.~Finn, and L.~E. Feldman, ``{Risk Prediction Model Versus United States Preventive Services Task Force Lung Cancer Screening Eligibility Criteria: Reducing Race Disparities.},'' {\em Journal of thoracic oncology : official publication of the International Association for the Study of Lung Cancer}, vol.~15, pp.~1738--1747, 11 2020.

\bibitem{tenHaaf2017RiskStudy}
K.~ten Haaf, J.~Jeon, M.~C. Tammem{\"{a}}gi, S.~S. Han, C.~Y. Kong, S.~K. Plevritis, E.~J. Feuer, H.~J. de~Koning, E.~W. Steyerberg, and R.~Meza, ``{Risk prediction models for selection of lung cancer screening candidates: A retrospective validation study},'' {\em PLOS Medicine}, vol.~14, no.~4, pp.~1--24, 2017.

\bibitem{Nobel1996HistogramPartitions}
A.~Nobel, ``{Histogram Regression Estimation Using Data-Dependent Partitions},'' {\em The Annals of Statistics}, vol.~24, pp.~1084--1105, 9 1996.

\bibitem{Mann1947OnOther}
H.~B. Mann and D.~R. Whitney, ``{On a Test of Whether one of Two Random Variables is Stochastically Larger than the Other},'' {\em Ann. Math. Statist.}, vol.~18, no.~1, pp.~50--60, 1947.

\bibitem{Shirahata1993EstimateStatistic}
S.~Shirahata, ``{Estimate of Variance of Wilcoxon-Mann-Whitney Statistic},'' {\em Journal of the Japanese Society of Computational Statistics}, vol.~6, no.~2, pp.~1--10, 1993.

\bibitem{Chipman2007MakingTrees}
H.~A. Chipman, E.~I. George, and R.~E. Mcculloch, ``{Making Sense of a Forest of Trees},'' in {\em Proceedings of the 30th Symposium on the Interface}, pp.~84--92, 2007.

\bibitem{Steingrimsson2021TransportingPopulation}
J.~A. Steingrimsson, C.~Gatsonis, and I.~J. Dahabreh, ``{Transporting a prediction model for use in a new target population},'' {\em American Journal of Epidemiology (In Press)}, 2021.

\bibitem{Pedregosa2011Scikit-learn:Python}
F.~Pedregosa, G.~Varoquaux, A.~Gramfort, V.~Michel, B.~Thirion, O.~Grisel, M.~Blondel, P.~Prettenhofer, R.~Weiss, V.~Dubourg, J.~Vanderplas, A.~Passos, D.~Cournapeau, M.~Brucher, M.~Perrot, and E.~Duchesnay, ``{Scikit-learn: Machine Learning in Python},'' {\em Journal of Machine Learning Research}, vol.~12, pp.~2825--2830, 2011.

\bibitem{Corbett-Davies2016AClear.}
S.~Corbett-Davies, E.~Pierson, A.~Feller, and S.~Goel, ``{A computer program used for bail and sentencing decisions was labeled biased against blacks. It’s actually not that clear.},'' 10 2016.

\bibitem{Angwin2016MachineBias}
J.~Angwin, J.~Larson, S.~Mattu, and L.~Kirchner, ``{Machine Bias},'' 5 2016.

\end{thebibliography}

\newpage 
\appendix 

\section{Details of the CART-TO algorithm}\label{sec:project2_appendix_cart-to-details}

More specifically, for a tree $T$ let $I_T$ denote the set of internal nodes. $|I_T|$ is the size of $I_T$ and thus $|I_T|+1$ is the number of terminal nodes. For each terminal node $l \in \{1, \ldots, |I_T|+1\}$, let $R_l\subseteq\mathcal{X}$ denote the subgroup it represents. For a given tuning parameter $\alpha \in \mathbb{R}^+$ define the cost-complexity of a tree $T$ as $\sum_{l=1}^{|I_T|+1}{\sum_{x_i\in R_l}{Q(\mu_i,\widehat{\mu}(R_l))}}+\alpha(|I_T|+1)$. The first term is the training error of the tree and the second term penalizes the complexity (size) of the tree where the size of the penalty term depends on $\alpha$. We say that a subtree of the initial tree $T_0$ is optimal for a fixed $\alpha$ if it minimizes the cost complexity. By varying $\alpha$ from $0$ to $\infty$ different subtrees of the initial tree $T_0$ become optimal and we denote the sequence of subtrees that are optimal w.r.t.~some value of $\alpha$ and the corresponding $\alpha$ values as $T_0, \ldots T_{\widehat K}$ and $\alpha_0, \ldots \alpha_{\widehat K}$. The last step is to use cross-validation to select the final tree from the sequence of candidate trees.

\begin{algorithm}
\caption{CART with transformed outcome}\label{alg:cart-to}
\KwIn{Data $\mathcal{D}=\{(X_i,Y_i):i=1,\dots,n\}$ and a prediction model $h:\mathcal{X}\to\mathcal{Y}$}
\KwOut{Subgroups and subgroup-specific estimators of model performance}
\nl Build a full-grown tree $T_0$\;
Start with the root tree that has only one node that contains all samples in $\mathcal{D}$\;
\While{there is a non-terminal node that has not been split into child nodes}{
\For{each of such node}{
\lIf{stopping conditions are met}{set node as terminal}
\lElse{split current node into two child nodes using the splitting criteria in expression~(\ref{eq:cart-to-splitting-criteria})}
}
}
\KwRet{the resulting tree as $T_0$}\;

\nl Prune fully grown tree\;
Start with $\alpha_0=0$\;
Increase $\alpha$ and find the subtree $T_m\subseteq T_0$ that minimizes $\sum_{l=1}^{|I_T|+1}{\sum_{x_i\in R_l}{Q(\mu_i,\widehat{\mu}(R_l))}}+\alpha(|I_T|+1)$\;
\KwRet{$\alpha_0<\alpha_1<\ldots<\alpha_{\widehat K}$ and a sequence of candidate trees $T_0\supset T_1\supset T_2\supset \ldots \supset T_{\widehat K}$}\;

\nl Select final subtree\;
For a fixed $V$, split $\mathcal{D}$ into mutually exclusive and exhaustive sets $\mathcal{D}_{1},\dots,\mathcal{D}_{V}$\;
\For{$v=1$ \KwTo $V$}{
build a fully grown tree $T_{v,0}$ as in Step 1 using all data but $\mathcal{D}_{v}$\;
\For{$k=0$ \KwTo $\widehat K$}{
calculate the model performance $mp(T_{v,k})$ of subtree $T_{v,k}$ of $T_{v,0}$ associated with $\alpha_k$\ using only data from $\mathcal{D}_v$.
}
}
Set $k^*\triangleq\argmin_{k\in\{0,1,\dots,K\}}{\sum_{v=1}^{V}{mp(T_{v,k})}}$\;
\nl For each terminal node calculate estimators of model performance using only data in the terminal node.
\\
\KwRet{$T_{k^*}$ as the final tree}\;
\end{algorithm}

\section{PASD algorithms}\label{sec:project2_appendix_PASD1-details}

\begin{algorithm}
\caption{PASD-1}\label{alg:PASD1}
\KwIn{Data $\mathcal{D}=\{(X_i,Y_i):i=1,\dots,n\}$, prediction model $h:\mathcal{X}\to\mathcal{Y}$}
\KwOut{A tree that partitions the covariate space}
\nl Build a full-grown tree $T_0$\;
Start with the null tree with only one node which contains all samples in $\mathcal{D}$\;
\While{there is a non-terminal node that has not been split into child nodes}{
\For{each such node}{
\lIf{stopping conditions are met}{set node as terminal}
\lElse{split current node into two child nodes with ($j^*$, $c^*$) from equation~(\ref{eq:PASD_split_criterion})}
}
}
\KwRet{the resulting tree as $T_0$}\;

\nl Prune $T_0$\;
Start with $\alpha_0=0$ and $i=0$\;
\While{$T_i$ contains more than the root node}{
set $i = i + 1$\;
set $m_{i-1}^*\triangleq \argmin_{m\in T_{i-1}}{g(m,T_{i-1})}$ and $\alpha_i\triangleq g(m_{i-1}^*,T_{i-1})$ with $g$ defined as in equation~(\ref{eq:g_func})\;
remove branch $T^{(m_{i-1}^*)}$ from $T_{i-1}$, and set resulting tree as $T_i$\;
}
\KwRet{$\alpha_0<\alpha_1<\ldots<\alpha_{\widehat K}$ with corresponding $T_0\supset T_1\supset T_2\supset \ldots \supset T_{\widehat K}$}\;

\nl Select final subtree\;
Split $\mathcal{D}$ into $\mathcal{D}_{1},\dots,\mathcal{D}_{V}$\;
\For{$v=1$ \KwTo $V$}{
build a fully grown tree $T_{v,0}$ as in step 1 using all data but $\mathcal{D}_{v}$\;
\For{$k=0$ \KwTo $K$}{
find the subtree $T_{v,k}$ of $T_{v,0}$ associated with $\alpha_k$\;
compute prediction mse $E(T_{v,k})=\text{avg}_i\{(\mu_i-\widehat{\mu}_i)^2\}$ using only data in $\mathcal{D}_{v}$\;
}
}
Set $k^*\triangleq\argmin_{k\in\{0,1,\dots,K\}}{\sum_{v=1}^{V}{E(T_{v,k})}}$\;
\KwRet{$T_{k^*}$ as the final subtree}\;

\end{algorithm}

\begin{algorithm}
\caption{PASD-2}\label{alg:project2_PASD2}
\KwIn{Data $\mathcal{D}=\{(X_i,Y_i):i=1,\dots,n\}$, prediction model $h:\mathcal{X}\to\mathcal{Y}$}
\KwOut{Subgroups and subgroup-specific estimators of model performance}
\nl Build a full-grown tree $T_0$\;
Start with the null tree with only one node which contains all samples in $\mathcal{D}$\;
\While{there is a non-terminal node that has not been split into child nodes}{
\For{each such node}{
\lIf{stopping conditions are met}{set node as terminal}
\lElse{split current node into two child nodes with ($j^*$, $c^*$) from equation~\eqref{eq:PASD_split_criterion}}
}
}
\KwRet{the resulting tree as $T_0$}\;

\nl Prune $T_0$\;
Start with $\alpha_0=0$ and $i=0$\;
\While{$T_i$ contains more than the root node}{
set $i = i + 1$\;
set $m_{i-1}^*\triangleq \argmin_{m\in T_{i-1}}{g(m,T_{i-1})}$ and $\alpha_i\triangleq g(m_{i-1}^*,T_{i-1})$ with $g$ defined as in equation~(\ref{eq:g_func})\;
remove branch $T^{(m_{i-1}^*)}$ from $T_{i-1}$, and set resulting tree as $T_i$\;
}
\KwRet{$\alpha_0<\alpha_1<\cdots<\alpha_{\widehat K}$ with corresponding $T_0\supset T_1\supset T_2\supset \cdots \supset T_{\widehat K}$}\;

\nl Select final subtree\;
Split $\mathcal{D}$ into $\mathcal{D}_{1},\dots,\mathcal{D}_{V}$\;
\For{$v=1$ \KwTo $V$}{
build a fully grown tree $T_{v,0}$ as in step 1 using all data but $\mathcal{D}_{v}$\;
\For{$k=0$ \KwTo $K$}{
find the subtree $T_{v,k}$ of $T_{v,0}$ associated with $\alpha_k$\;
compute $S_{\alpha^\prime}(T_{v,k})$ using only observations in $\mathcal{D}_{v}$\;
}
}
Set $k^*\triangleq\argmax_{k\in\{0,1,\dots,\widehat K\}}{\sum_{v=1}^{V}{S_{\alpha^\prime}(T_{v,k})}}$\;
\KwRet{$T_{k^*}$ as the final tree}\;
\end{algorithm}

\newpage
\section{Proof of Theorem 1}\label{sec:project2_appendix_thm1_proof}

Following similar notations from \cite{Nobel1996HistogramPartitions}, let $\widehat{\psi}_n(\mathcal{D}_n)$ be the resulting partition of $\mathcal{X}$ when we apply the partition rule $\widehat{\psi}_n$ to $\mathcal{D}_n$. Denote $\widehat{\psi}_n[x]$ as the unique cell of $\widehat{\psi}_n(\mathcal{D}_n)$ containing the vector $x$. Let $\Pi_n$ be the range of $\widehat{\psi}_n$.

\begin{proof}
Define a transformed outcome $Y^*=l(Y,h(X))$. The tree estimator for model performance $\mu(x)=\mathbb{E}[l(Y,h(X))|X=x]=\mathbb{E}(Y^*|X=x)$ can be written as
\begin{equation}
    \widehat{\mu}_n(x)=\frac{\sum_{i=1}^{n}{I(X_i\in\widehat{\psi}_n[x])\cdot Y_i^*}}{\sum_{i=1}^{n}{I(X_i\in\widehat{\psi}_n[x])}}.
\end{equation}
Further, define
\begin{equation}
    \tilde{\mu}_n(x)=\frac{\sum_{i=1}^{n}{I(X_i\in\widehat{\psi}_n[x])\cdot \mu(X_i)}}{\sum_{i=1}^{n}{I(X_i\in\widehat{\psi}_n[x])}}.
\end{equation}
Since
\begin{equation}
    |\mu(x)-\widehat{\mu}_n(x)|^2 \le 2|\mu(x)-\tilde{\mu}_n(x)|^2+2|\tilde{\mu}_n(x)-\widehat{\mu}_n(x)|^2, \quad\forall x\in\mathcal{X},
\end{equation}
we have
\begin{equation}
    \int{|\mu(x)-\widehat{\mu}_n(x)|^2}\:\mathrm{d}{P(x)}\le \int{2|\mu(x)-\tilde{\mu}_n(x)|^2}\:\mathrm{d}{P(x)}+\int{2|\tilde{\mu}_n(x)-\widehat{\mu}_n(x)|^2}\:\mathrm{d}{P(x)}, \label{proof_ineq}
\end{equation}
where $P$ is the distribution function of $X$. We next show both terms on the RHS of (\ref{proof_ineq}) are $o_p(1)$. 

By definitions of $\tilde{\mu}_n$ and $\widehat{\mu}_n$, and equation (18) in \cite{Nobel1996HistogramPartitions},
\begin{align}
    &\int{|\tilde{\mu}_n(x)-\widehat{\mu}_n(x)|^2}\:\mathrm{d}{P(x)}\\
    &=\int{\left|\frac{\sum_{i=1}^{n}{I(X_i\in\widehat{\psi}_n[x])\cdot [Y_i^*-\mu(X_i)]}}{\sum_{i=1}^{n}{I(X_i\in\widehat{\psi}_n[x])}}\right|^2}\:\mathrm{d}{P(x)} \\
    &=\sum_{A\in\widehat{\psi}_n}{\int_{A}{\left|\frac{\sum_{i=1}^{n}{I(X_i\in A)\cdot[Y_i^*-\mu(X_i)]}}{\sum_{i=1}^{n}{I(X_i\in A)}}\right|^2}\:\mathrm{d}{P(x)}} \\
    &=\sum_{A\in\widehat{\psi}_n}{\left|\frac{\sum_{X_i\in A}{[Y_i^*-\mu(X_i)]}}{n_A}\right|^2\cdot P(A)} \\
    &\le 2B\sum_{A\in\widehat{\psi}_n}{\left|\frac{\sum_{X_i\in A}{[Y_i^*-\mu(X_i)]}}{n_A}\right|\cdot P(A)} \label{ineq_bound1}\\
    &=2B\sum_{A\in\widehat{\psi}_n}{\left|\frac{\sum_{X_i\in A}{[Y_i^*-\mu(X_i)]}}{n_A}\right|\cdot \widehat{P}_n(A)}+2B\sum_{A\in\widehat{\psi}_n}{\left|\frac{\sum_{X_i\in A}{[Y_i^*-\mu(X_i)]}}{n_A}\right|\cdot[P(A)-\widehat{P}_n(A)]}\\
    &\le 2B\sum_{A\in\widehat{\psi}_n}{\left|\frac{\sum_{X_i\in A}{[Y_i^*-\mu(X_i)]}}{n_A}\right|\cdot\widehat{P}_n(A)}+4B^2\sum_{A\in\widehat{\psi}_n}{|P(A)-\widehat{P}_n(A)|}\label{ineq_bound2}\\
    &=2B\sum_{A\in\widehat{\psi}_n}{\left|\frac{\sum_{X_i\in A}{[Y_i^*-\mu(X_i)]}}{n}\right|} + o_p(1) \label{eq_deviation_bound}\\
    &\le2B\sup_{\pi\in\Pi_n}{\sum_{A\in\pi}{\left|\frac{\sum_{X_i\in A}{[Y_i^*-\mu(X_i)]}}{n}\right|}}+o_p(1).
\end{align}
where $n_A$ is the number of observations in cell $A$, $P(A)=\int_{A}\:\mathrm{d}P(x)$, $\widehat{P}_n(A)=\frac{1}{n}\sum_{i=1}^{n}{I(X_i\in A)}=\frac{n_A}{n}$, and, by assumption, $\max\{Y^*,\mu(X)\}\le B<\infty$. Inequalities (\ref{ineq_bound1}) and (\ref{ineq_bound2}) follow from the bounded assumption of $Y$ and $h$. Equation (\ref{eq_deviation_bound}) follows from equation (17) of \cite{Nobel1996HistogramPartitions}.

Define $f(x,y)=y^*-\mu(x)$ for $(x,y)\in\mathcal{X}\times[-B,B]$. $f$ is thus bounded and $\int_{A\times[-B,B]}{f(x,y)}\:\mathrm{d}{Q(x,y)}=0$ for every measurable subset $A\in\mathcal{X}$, where $Q$ is the joint distribution of $(X,Y)$. Therefore, 
\begin{align}
    &\int{|\tilde{\mu}_n(x)-\widehat{\mu}_n(x)|^2}\:\mathrm{d}{P(x)}\\
    &\le2B\sup_{\pi\in\Pi_n}{\sum_{A\in\pi}{\left|\frac{\sum_{X_i\in A}{[Y_i^*-\mu(X_i)]}}{n}\right|}}+o_p(1)\\
    &=2B\sup_{\tilde{\pi}\in\tilde{\Pi}_n}{\sum_{V\in\tilde{\pi}}{\left|\frac{\sum_{(X_i,Y_i)\in V}{f(X_i,Y_i)}}{n}-\int_{V}{f(x,y)}\:\mathrm{d}{Q(x,y)}\right|}}+o_p(1)\\
    &=o_p(1),
\end{align}
where for each $\pi\in\Pi_n$, $\tilde{\pi}=\{A_i\times[-B,B]:A_i\in\pi\}$ is the associated partition of $\mathcal{X}\times[-B,B]$, and $\tilde{\Pi}_n=\{\tilde{\pi}_n:\pi\in\Pi_n\}$. The last equality follows from Proposition 2, Lemma 3 and equation (13) in \cite{Nobel1996HistogramPartitions}. 

Following the same argument as in the proof of Theorem 1 in \cite{Nobel1996HistogramPartitions}, one can also show $\int{|\mu(x)-\tilde{\mu}_n(x)|^2}\:\mathrm{d}{P(x)}=o_p(1)$. Hence $\int{|\mu(x)-\widehat{\mu}_n(x)|^2}\:\mathrm{d}{P(x)}=o_p(1)$ and $\widehat{\mu}_n$ is $L_2$-consistent. 

\end{proof}

\section{Estimation of model performance and its variance estimator} \label{sec:project2_appendix_acc_estimation}

The PASD tree algorithm depends on estimating the model performance $\mu(w)$ and the variance of its estimator $\widehat{\text{Var}}[\widehat{\mu}(w)]$. In this section, we present the calculation for loss-based model performance measures (individual-level) and AUC (group-level).

\subsection{Loss-based measures}

For individual-level loss-based model performance measure $\mu=L(Y,h(X))$, $\mu(w)=\mathbb{E}[L(Y,h(X))\mid X\in w]$ can be estimated by the sample mean
\begin{equation}
\widehat{\mu}(w) = \frac{1}{n_w}\sum_{i:\:X_i\in w}{L(Y_i,h(X_i))}=\frac{1}{n_w}\sum_{i:X_i\in w}{\mu_i}
\end{equation} 
which is unbiased and consistent. $n_w$ is the number of observations with covariate vector $X\in w$. The variance of $\widehat{\mu}(w)$ can then be estimated by 
\begin{equation}
\widehat{\text{Var}}[\widehat{\mu}(w)] = \frac{1}{n_w(n_w-1)}\sum_{i:\: X_i\in w}{[\mu_i-\widehat{\mu}(w)]^2}.
\end{equation}
$\widehat{\text{Var}}[\widehat{\mu}(w)]$ is unbiased and consistent for $\text{Var}[\widehat{\mu}(w)]$.

\subsection{Binary Outcome with AUC}
For binary response and AUC as the model performance measure, $\mu(w)=\mathbb{E}[I\{\widehat{r}(X)>\widehat{r}(X^\prime)\}\mid X\in w, X^\prime\in w, Y=1,Y^\prime=0]$ can be estimated by the Mann-Whitney U statistic \cite{Mann1947OnOther}
\begin{equation}
\widehat{\mu}(w) \triangleq \frac{1}{n_{w_0}\cdot n_{w_1}}\sum_{i:\:X_i\in w,Y_i=1}{\sum_{j:\:X_j\in w,Y_j=0}{I\{\widehat{r}(X_i)>\widehat{r}(X_j)\}}} \label{mann_whitney_u}
\end{equation}
where $n_{w_1}$ is the number of observations with covariate vector in $w$ and outcome equal to 1. Similarly for $n_{w_0}$. $\widehat{\mu}(w)$ is the minimum variance unbiased estimator of $\mu(w)$ since it is a U-statistic. Its variance can be found using the standard two-sample U-statistic variance formula 
\begin{equation}
\widehat{\mu}(w) \triangleq \frac{1}{n_{w_0}\cdot n_{w_1}}\sum_{i:\:X_i\in w,Y_i=1}{\sum_{j:\:X_j\in w,Y_j=0}{I\{\widehat{r}(X_i)>\widehat{r}(X_j)\}}}
\end{equation}
can be found through the standard two-sample U-statistics formula. 
\begin{equation}
\text{Var}[\widehat{\mu}(w)] = \frac{\mu(w)[1-\mu(w)]+(n_{w_1}-1)\xi_{w,0,1}+(n_{w_0}-1)\xi_{w,1,0}}{n_{w_0}\cdot n_{w_1}}
\end{equation}
where
\begin{align}
&\xi_{w,0,1} \notag\\
&= \mathbb{E}[I\{\widehat{r}(X^\prime)>\widehat{r}(X),\widehat{r}(X^{\prime\prime})>\widehat{r}(X)\}\mid Y^\prime=1,Y^{\prime\prime}=1,Y=0,X^\prime\in w,X^{\prime\prime}\in w,X\in w] \notag\\
&\quad - \mu(w)^2
\end{align}
and 
\begin{align}
&\xi_{w,1,0} \notag\\
&= \mathbb{E}[I\{\widehat{r}(X)>\widehat{r}(X^\prime),\widehat{r}(X)>\widehat{r}(X^{\prime\prime})\}\mid Y^\prime=0,Y^{\prime\prime}=0,Y=1,X^\prime\in w,X^{\prime\prime}\in w,X\in w] \notag\\
&\quad - \mu(w)^2.
\end{align}
An unbiased estimator of $\text{Var}[\widehat{\mu}(w)]$ can be obtained by finding unbiased estimators for $\mu(w)$, $\mu(w)^2$, $\xi_{w,0,1}$ and $\xi_{w,1,0}$ separately, which are given by
\begin{itemize}
\item unbiased estimator for $\mu(w)$ is the $\widehat{\mu}(w)$ defined in (\ref{mann_whitney_u}).
\item unbiased estimator for $\mu(w)^2$ is
\begin{equation}
\widehat{\mu(w)^2} = \frac{\sum_{i,j,k,l}{I\{\widehat{r}(X_i)>\widehat{r}(X_j),\widehat{r}(X_k)>\widehat{r}(X_l)\}}}{n_{w_1}(n_{w_1}-1)n_{w_0}(n_{w_0}-1)}
\end{equation}
with summation over $(i,j,k,l)\in\{(i,j,k,l): i\neq k,j\neq l, X_i\in w,X_j\in w,X_k\in w,X_l\in w, Y_i=1,Y_k=1,Y_j=0,Y_l=0\}$.
\item unbiased estimator for $\xi_{w,0,1}$ is
\begin{equation}
\widehat{\xi}_{w,0,1} = \frac{\sum_{i,j,k}{I\{\widehat{r}(X_i)>\widehat{r}(X_k),\widehat{r}(X_j)>\widehat{r}(X_k)\}}}{n_{w_1}(n_{w_1}-1)n_{w_0}} - \widehat{\mu(w)^2}
\end{equation}
with summation over $(i,j,k)\in\{(i,j,k): i\neq j, X_i\in w,X_j\in w,X_k\in w, Y_i=1,Y_j=1,Y_k=0\}$. 
\item unbiased estimator for $\xi_{w,1,0}$ is
\begin{equation}
\widehat{\xi}_{w,1,0} = \frac{\sum_{i,j,k}{I\{\widehat{r}(X_i)>\widehat{r}(X_j),\widehat{r}(X_i)>\widehat{r}(X_k)\}}}{n_{w_1}n_{w_0}(n_{w_0}-1)} - \widehat{\mu(w)^2}
\end{equation}
with summation over $(i,j,k)\in\{(i,j,k): j\neq k, X_i\in w,X_j\in w,X_k\in w, Y_i=1,Y_j=0,Y_k=0\}$. 
\end{itemize}
In fact, the above 4 estimators are all U-statistics. Finally, the unbiased estimator for $\text{Var}[\widehat{\mu}(w)]$ is
\begin{equation}
\widehat{\text{Var}}[\widehat{\mu}(w)] 
= \frac{\widehat{\mu}(w) - \widehat{\mu(w)^2}+(n_{w_1}-1)\widehat{\xi}_{w,0,1}+(n_{w_0}-1)\widehat{\xi}_{w,1,0}}{n_{w_0}\cdot n_{w_1}}.
\end{equation}
\cite{Shirahata1993EstimateStatistic} argued that $\widehat{\text{Var}}[\widehat{\mu}(w)]$ is the UMVU estimator for $\text{Var}[\widehat{\mu}(w)]$.

\section{Random forest and gradient boosting algorithm}
\label{app:Sec-ensemble}
Algorithm \ref{alg:random_PASD_forest} gives the random forest algorithm with PASD trees and Algorithm \ref{alg:gb_PASD} presents the gradient boosting algorithm with PASD trees.

\begin{algorithm}
\caption{Random PASD Forest}\label{alg:random_PASD_forest}
\nl Generate a bootstrap sample of the training data $D$\;
\nl On each bootstrap sample, fit a fully grown PASD tree model where only a random subset of covariates is considered each time a splitting decision is made. Calculate the terminal node estimators using the out-of-bag sample\; 
\nl The final prediction of the random PASD forest is given by the average prediction of the $B$ PASD trees from Step 2.
\end{algorithm}

\begin{algorithm}
\caption{Gradient Boosting PASD Trees}\label{alg:gb_PASD}
\nl Initialize $T_0(x)=\frac{1}{n}\sum_{i=1}^{n}{\mu(x_i)}$\;
\nl \For{m=1,\dots,M}{
\begin{enumerate}
    \item compute residuals
    \begin{equation*}
        r_{i} = \mu(x_i)-T_{m-1}(x_i)
    \end{equation*}
    \item fit a PASD tree $\widetilde{T}_m$ to data $\{(x_i,r_i):i=1,\dots,n\}$
    \item set 
    \begin{equation*}
        T_m(x) = T_{m-1}(x) + \lambda\cdot\widetilde{T}_m(x)
    \end{equation*}
\end{enumerate}
}
\nl Output $T_M(x)$;
\end{algorithm}

\section{Algorithms for model combination}
\label{app:Model-comb}

Algorithm \ref{alg:project3_mv_combination} gives the algorithm for combining prediction models using majority voting.

\begin{algorithm}
\caption{Majority voting guided model combination}\label{alg:project3_mv_combination}
\For{$b=1$ \KwTo $B$}{
Generate a bootstrap sample $D^{(b)}$ from $D$\;
\For{$k=1$ \KwTo $K$}{
    Fit PASD tree $\widehat{\mu}_{k}^{(b)}$ with $D^{(b)}$ and $h_k$ as in Algorithm~\ref{alg:PASD1} Step 1 but only consider a random subset of covariates in each stage of splitting\;
    Estimate the terminal nodes model performance with the OOB sample\;
    }
} 
\KwRet{$\widehat{\pi}_{k\mid D}(x)=\frac{1}{B}\sum_{b=1}^{B}{I\left\{\widehat{\mu}_{k}^{(b)}(x) = \min_{t}{\widehat{\mu}_{t}^{(b)}(x)}\right\}}$}\;
\end{algorithm}

\section{Additional simulation results}
\label{app:sim}

\subsection{True underlying tree structure for each simulation setting}
\label{sec:project2_appendix_true_tree_structure}

In this section, we present the true tree structure in each simulation setting. Note that in setting 2 and 3, if a tree first splits on $X_5$ and then $X_3$, it is also considered as correct as long as the same four subgroups are identified.

\begin{figure}[!htb]
        \centering
        \includegraphics[scale=0.5]{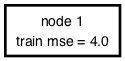}
        \caption{True underlying tree structure for simulation setting 1.}
        \label{img:sim_case1_true_tree}
    \end{figure}

\begin{figure}[!htb]
        \centering
        \includegraphics[scale=0.5]{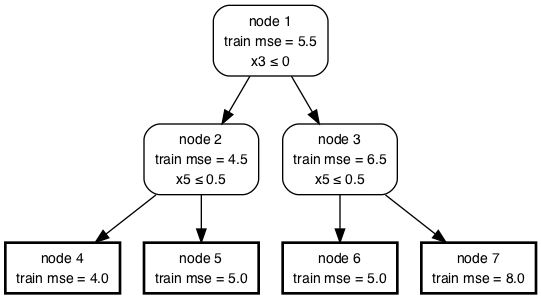}
        \caption{True underlying tree structure for simulation setting 2.}
        \label{img:sim_case2_true_tree}
    \end{figure}

\begin{figure}[!htb]
        \centering
        \includegraphics[scale=0.5]{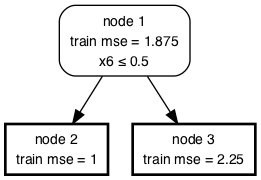}
        \caption{True underlying tree structure for simulation setting 4.}
        \label{img:sim_case4_true_tree}
    \end{figure}

\subsection{Evaluation metrics}
\label{app:eval-sim}

To evaluate the fitted trees we use a dataset of size 1000 sampled from the same distribution as the dataset used to build the tree that is independent of the data used for tree building. To compare the three algorithms we use the following metrics:
\begin{enumerate}
\item \textit{Mean squared error (MSE)}. Running each observation down the fitted tree gives a prediction $\widehat \mu_i$ of model performance for each observation $\mu_i$. The mean squared error is defined as $\frac{1}{N}\sum_{i=1}^N {(\mu_i-\widehat{\mu}_i)^2}$, where $\mu_i$ is the observed model performance for observation $i$. This measures the ability of the tree algorithms to estimate model performance for future observations. 

\item Pairwise prediction similarity (PPS) between the true tree and the fitted tree \cite{Chipman2007MakingTrees}. Let $I_T(i,j)$ and $I_F(i,j)$ be the indicators of whether observations $i$ and $j$ are placed in the same subgroup (terminal node) in the true tree and the fitted tree, respectively. PPS is defined as
\begin{equation}
    PPS = 1 - \frac{\sum_{i>j}{|I_T(i,j)-I_F(i,j)|}}{{\binom{N}{2}}}.
\end{equation}
It is the proportion of observations that get classified in the same subgroup by the true and the fitted tree and higher values imply better concordance between the true and fitted tree.

\item Proportion of fitted trees that only split on the covariates that affect prediction performance (e.g.,~prediction error modifiers \cite{Steingrimsson2021TransportingPopulation}). In simulation setting 1 there are no prediction error modifiers. For simulation settings 2 and 3, $X^{(3)}$ and $X^{(5)}$ are prediction error modifiers, and for simulation setting 4, $X^{(6)}$ is a prediction error modifier. 

\item Proportion of fitted trees (out of 1,000) that only split on the correct covariates and have an equal number of subgroups compared to the true tree. For setting 1, there should be zero splits (one terminal node). We expect three splits for settings 2 and 3, and one split for setting 4.

\end{enumerate}

One thing to note is that the calculations of prediction MSE and pairwise prediction dissimilarity only require knowing the subgroups (terminal nodes) defined by the true tree and the fitted PASD tree rather than the structure of the trees. In fact, there can be multiple true trees as long as they produce the same terminal nodes. For example, in simulation setting 2, the true tree can first split on $X^{(5)}$ and then $X^{(3)}$, or it can split on $X^{(3)}$ then followed by $X^{(5)}$. Both trees are considered correct as long as they produce the same four subgroups; and in this case, calculations of prediction MSE and pairwise prediction dissimilarity are not affected.

\subsection{Additional Simulation Results}\label{sec:project2_appendix_additional_sim_results}

In addition to evaluating the fitted trees by prediction performance, we compare how often the fitted trees overfit/underfit the data. The tree algorithms overfit when they discover subgroups that do not exist. These errors can arise either when the fitted tree is splitting on noise variables or it has the correct variables but is splitting too many times. Underfitting occurs when the tree fails to discover a true subgroup, and we define it as the tree having no noise variables but not enough splits. In particular, we examine 1) the set of variables that the tree is splitting on, and 2) the number of subgroups (terminal nodes) in the fitted tree to determine what type of error the tree has. Let $S_F$ and $S_T$ denote the set of covariates used in the fitted tree and the set of correct (expected) covariates in the true tree, respectively. Let $n_F$ and $n_T$ denote the number of subgroups in the fitted tree and the true tree, respectively. 
\begin{enumerate}
   \item[(0)] $S_F \subset S_T$. For example, $S_F=\{X^{(3)}\}$ and $S_T=\{X^{(3)},X^{(5)}\}$.
         \begin{enumerate}
             \item[(0)] $n_F<n_T$. This is underfitting. 
             \item[(1)] $n_F=n_T$. This is overfitting.
             \item[(2)] $n_F>n_T$. This is overfitting. 
         \end{enumerate}
   \item[(1)] $S_F = S_T$. For example, $S_F=S_T=\{X^{(3)},X^{(5)}\}$.
         \begin{enumerate}
             \item[(0)] $n_F<n_T$. This is underfitting. 
             \item[(1)] $n_F=n_T$. This is good fit with no error.
             \item[(2)] $n_F>n_T$. This is overfitting. 
         \end{enumerate}
   \item[(2)] $S_F \supset S_T$. For example, $S_F=\{X^{(1)},X^{(3)},X^{(5)}\}$ and $S_T=\{X^{(3)},X^{(5)}\}$.
         \begin{enumerate}
             \item[(*)] $n_F<n_T$ or $n_F=n_T$ or $n_F>n_T$. This is overfitting. 
         \end{enumerate}
\end{enumerate}

\begin{table}[!htb]
\centering
\caption{Results of the four simulation settings with $\alpha^\prime=2$. }
\label{tab:project2_sim_results_alphaPrime_2}
\resizebox{\textwidth}{!}{%
\begin{tabular}{@{}ccccccccccccc@{}}
\toprule
\multirow{2}{*}{setting} &
  \multirow{2}{*}{model} &
  \multirow{2}{*}{mean mse} &
  \multirow{2}{*}{mean pps} &
  \multicolumn{5}{c}{overfitting} &
  \multicolumn{3}{c}{underfitting} &
  \multirow{2}{*}{good fit} \\ \cmidrule(lr){5-12}
  &       &        &       & total & (2,.) & (1,2) & (0,1) & (0,2) & total & (1,0) & (0,0) &       \\ \midrule
1 & cart  & 32.095 & 0.976 & 0.048 & 0.048 & 0     & 0     & 0     & 0     & 0     & 0     & 0.952 \\
1 & PASD1 & 32.110 & 0.961 & 0.085 & 0.085 & 0     & 0     & 0     & 0     & 0     & 0     & 0.915 \\
1 & PASD2 & 32.126 & 0.934 & 0.123 & 0.123 & 0     & 0     & 0     & 0     & 0     & 0     & 0.877 \\
2 & cart  & 57.077 & 0.765 & 0.308 & 0.298 & 0.01  & 0     & 0     & 0.649 & 0.439 & 0.21  & 0.043 \\
2 & PASD1 & 57.134 & 0.780 & 0.479 & 0.455 & 0.024 & 0     & 0     & 0.447 & 0.254 & 0.193 & 0.074 \\
2 & PASD2 & 57.124 & 0.766 & 0.431 & 0.404 & 0.027 & 0     & 0     & 0.483 & 0.257 & 0.226 & 0.086 \\
3 & cart  & 56.647 & 0.769 & 0.325 & 0.316 & 0.008 & 0     & 0.001 & 0.638 & 0.435 & 0.203 & 0.037 \\
3 & PASD1 & 56.731 & 0.772 & 0.498 & 0.481 & 0.017 & 0     & 0     & 0.43  & 0.217 & 0.213 & 0.072 \\
3 & PASD2 & 56.708 & 0.768 & 0.474 & 0.461 & 0.013 & 0     & 0     & 0.452 & 0.238 & 0.214 & 0.074 \\
4 & cart  & 7.950  & 0.977 & 0.087 & 0.087 & 0     & 0     & 0     & 0     & 0     & 0     & 0.913 \\
4 & PASD1 & 7.953  & 0.968 & 0.136 & 0.136 & 0     & 0     & 0     & 0     & 0     & 0     & 0.864 \\
4 & PASD2 & 7.954  & 0.961 & 0.156 & 0.156 & 0     & 0     & 0     & 0     & 0     & 0     & 0.844 \\ \bottomrule
\end{tabular}%
}
\end{table}

\begin{table}[!htb]
\centering
\caption{Results of the four simulation settings with $\alpha^\prime=3$. }
\label{tab:project2_sim_results_alphaPrime_3}
\resizebox{\textwidth}{!}{%
\begin{tabular}{@{}ccccccccccccc@{}}
\toprule
\multirow{2}{*}{setting} &
  \multirow{2}{*}{model} &
  \multirow{2}{*}{mean mse} &
  \multirow{2}{*}{mean pps} &
  \multicolumn{5}{c}{overfitting} &
  \multicolumn{3}{c}{underfitting} &
  \multirow{2}{*}{good fit} \\ \cmidrule(lr){5-12}
  &       &        &       & total & (2,.) & (1,2) & (0,1) & (0,2) & total & (1,0) & (0,0) &       \\ \midrule
1 & cart  & 32.095 & 0.976 & 0.048 & 0.048 & 0     & 0     & 0     & 0     & 0     & 0     & 0.952 \\
1 & PASD1 & 32.110 & 0.961 & 0.085 & 0.085 & 0     & 0     & 0     & 0     & 0     & 0     & 0.915 \\
1 & PASD2 & 32.081 & 0.983 & 0.041 & 0.041 & 0     & 0     & 0     & 0     & 0     & 0     & 0.959 \\
2 & cart  & 57.077 & 0.765 & 0.308 & 0.298 & 0.01  & 0     & 0     & 0.649 & 0.439 & 0.21  & 0.043 \\
2 & PASD1 & 57.134 & 0.780 & 0.479 & 0.455 & 0.024 & 0     & 0     & 0.447 & 0.254 & 0.193 & 0.074 \\
2 & PASD2 & 57.346 & 0.638 & 0.163 & 0.154 & 0.009 & 0     & 0     & 0.775 & 0.298 & 0.477 & 0.062 \\
3 & cart  & 56.647 & 0.769 & 0.325 & 0.316 & 0.008 & 0     & 0.001 & 0.638 & 0.435 & 0.203 & 0.037 \\
3 & PASD1 & 56.731 & 0.772 & 0.498 & 0.481 & 0.017 & 0     & 0     & 0.43  & 0.217 & 0.213 & 0.072 \\
3 & PASD2 & 56.852 & 0.640 & 0.21  & 0.205 & 0.005 & 0     & 0     & 0.746 & 0.27  & 0.476 & 0.044 \\
4 & cart  & 7.950  & 0.977 & 0.087 & 0.087 & 0     & 0     & 0     & 0     & 0     & 0     & 0.913 \\
4 & PASD1 & 7.953  & 0.968 & 0.136 & 0.136 & 0     & 0     & 0     & 0     & 0     & 0     & 0.864 \\
4 & PASD2 & 7.939  & 0.990 & 0.06  & 0.06  & 0     & 0     & 0     & 0     & 0     & 0     & 0.94  \\ \bottomrule
\end{tabular}%
}
\end{table}

\begin{table}[!htb]
\centering
\caption{Results of the four simulation settings with $\alpha^\prime=\log(n)$. }
\label{tab:sim_results}
\resizebox{\textwidth}{!}{%
\begin{tabular}{@{}ccccccccccccc@{}}
\toprule
\multirow{2}{*}{setting} &
  \multirow{2}{*}{model} &
  \multirow{2}{*}{mean mse} &
  \multirow{2}{*}{mean pps} &
  \multicolumn{5}{c}{overfitting} &
  \multicolumn{3}{c}{underfitting} &
  \multirow{2}{*}{good fit} \\ \cmidrule(lr){5-12}
  &       &        &       & total & (2,.) & (1,2) & (0,1) & (0,2) & total & (1,0) & (0,0) &       \\ \midrule
1 & cart  & 32.095 & 0.976 & 0.048 & 0.048 & 0     & 0     & 0     & 0     & 0     & 0     & 0.952 \\
1 & PASD1 & 32.110 & 0.961 & 0.085 & 0.085 & 0     & 0     & 0     & 0     & 0     & 0     & 0.915 \\
1 & PASD2 & 32.069 & 0.994 & 0.017 & 0.017 & 0     & 0     & 0     & 0     & 0     & 0     & 0.983 \\
2 & cart  & 57.077 & 0.765 & 0.308 & 0.298 & 0.010 & 0     & 0     & 0.649 & 0.439 & 0.210 & 0.043 \\
2 & PASD1 & 57.134 & 0.780 & 0.479 & 0.455 & 0.024 & 0     & 0     & 0.447 & 0.254 & 0.193 & 0.074 \\
2 & PASD2 & 57.957 & 0.372 & 0.021 & 0.020 & 0.001 & 0     & 0     & 0.975 & 0.086 & 0.889 & 0.004 \\
3 & cart  & 56.647 & 0.769 & 0.325 & 0.316 & 0.008 & 0     & 0     & 0.638 & 0.435 & 0.203 & 0.037 \\
3 & PASD1 & 56.731 & 0.772 & 0.498 & 0.481 & 0.017 & 0     & 0     & 0.430 & 0.217 & 0.213 & 0.072 \\
3 & PASD2 & 57.345 & 0.369 & 0.017 & 0.017 & 0.000 & 0     & 0     & 0.977 & 0.082 & 0.895 & 0.006 \\
4 & cart  & 7.950  & 0.977 & 0.087 & 0.087 & 0     & 0     & 0     & 0     & 0     & 0     & 0.913 \\
4 & PASD1 & 7.953  & 0.968 & 0.136 & 0.136 & 0     & 0     & 0     & 0     & 0     & 0     & 0.864 \\
4 & PASD2 & 7.936  & 0.995 & 0.034 & 0.034 & 0     & 0     & 0     & 0.001 & 0     & 0.001 & 0.965 \\ \bottomrule
\end{tabular}%
}
\end{table}

\begin{table}[!thb]
\centering
\caption{Results of the four simulation settings with $\alpha^\prime=4$. }
\label{tab:sim_results_alphaPrime_4}
\resizebox{\textwidth}{!}{%
\begin{tabular}{@{}ccccccccccccc@{}}
\toprule
\multirow{2}{*}{setting} &
  \multirow{2}{*}{model} &
  \multirow{2}{*}{mean mse} &
  \multirow{2}{*}{mean pps} &
  \multicolumn{5}{c}{overfitting} &
  \multicolumn{3}{c}{underfitting} &
  \multirow{2}{*}{good fit} \\ \cmidrule(lr){5-12}
  &       &        &       & total & (2,.) & (1,2) & (0,1) & (0,2) & total & (1,0) & (0,0) &       \\ \midrule
1 & cart  & 32.095 & 0.976 & 0.048 & 0.048 & 0     & 0     & 0     & 0     & 0     & 0     & 0.952 \\
1 & PASD1 & 32.110 & 0.961 & 0.085 & 0.085 & 0     & 0     & 0     & 0     & 0     & 0     & 0.915 \\
1 & PASD2 & 32.071 & 0.993 & 0.022 & 0.022 & 0     & 0     & 0     & 0     & 0     & 0     & 0.978 \\
2 & cart  & 57.077 & 0.765 & 0.308 & 0.298 & 0.01  & 0     & 0     & 0.649 & 0.439 & 0.21  & 0.043 \\
2 & PASD1 & 57.134 & 0.780 & 0.479 & 0.455 & 0.024 & 0     & 0     & 0.447 & 0.254 & 0.193 & 0.074 \\
2 & PASD2 & 57.681 & 0.494 & 0.055 & 0.052 & 0.003 & 0     & 0     & 0.913 & 0.197 & 0.716 & 0.032 \\
3 & cart  & 56.647 & 0.769 & 0.325 & 0.316 & 0.008 & 0     & 0.001 & 0.638 & 0.435 & 0.203 & 0.037 \\
3 & PASD1 & 56.731 & 0.772 & 0.498 & 0.481 & 0.017 & 0     & 0     & 0.43  & 0.217 & 0.213 & 0.072 \\
3 & PASD2 & 57.086 & 0.510 & 0.076 & 0.075 & 0.001 & 0     & 0     & 0.896 & 0.207 & 0.689 & 0.028 \\
4 & cart  & 7.950  & 0.977 & 0.087 & 0.087 & 0     & 0     & 0     & 0     & 0     & 0     & 0.913 \\
4 & PASD1 & 7.953  & 0.968 & 0.136 & 0.136 & 0     & 0     & 0     & 0     & 0     & 0     & 0.864 \\
4 & PASD2 & 7.937  & 0.993 & 0.043 & 0.043 & 0.0   & 0.0   & 0.0   & 0.0   & 0.0   & 0.0   & 0.957 \\ \bottomrule
\end{tabular}%
}
\end{table}

\section{Details of the EM algorithms}\label{sec:project3_appendix_em_details}

\subsection{Derivation of Algorithm~\ref{alg:project3_numerical_em}}
We use a latent variable representation by defining latent variables $Z_i$ such that 
\begin{equation}
    Y_i\mid X_i,Z_i=k\sim N(h_k(X_i),\sigma_k^2),
\end{equation}
and $P(Z_i=k\mid X_i)=\pi_k(X_i;\nu)$.

The complete-data likelihood is 
\begin{align}
    & \quad p(y,z\mid x;\gamma) \nonumber\\
    &= \prod_{i=1}^{n}{p(y_i\mid x_i,z_i;\gamma)\cdot p(z_i\mid x_i;\gamma)} \nonumber\\
    &= \prod_{i=1}^{n}{\left[\prod_{k=1}^{K}{\phi(y_i;h_k(x_i),\sigma_k^2)^{I(z_i=k)}}\right] \cdot \left[\prod_{k=1}^{K}{\pi_k(x_i;\nu)^{I(z_i=k)}}\right] }. \nonumber
\end{align}

The complete-data log-likelihood is thus
\begin{align}
    &\quad \log p(y,z\mid x;\gamma) \nonumber\\
    &= \sum_{i=1}^{n}{\sum_{k=1}^{K}{I(z_i=k)\left[ \log \pi_k(x_i;\nu) + \log \phi(y_i;h_k(x_i),\sigma_k^2)\right]}}.
\end{align}

Marginalize out the hidden variables $Z$ by taking the expectation of the complete-data log-likelihood w.r.t. $p(z\mid y,x;\gamma^\prime)$ where $\gamma^\prime$ is some value of the parameter. This gives
\begin{align}
    &\quad Q(\gamma\mid\gamma^\prime)\nonumber\\
    &= \mathbb{E}\left[\log p(y,Z\mid x;\gamma)\mid y,x;\gamma^\prime\right] \nonumber\\
    &=\sum_{i=1}^{n}{\sum_{k=1}^{K}{\underbrace{P(Z_i=k\mid y_i,x_i;\gamma^\prime)}_{\lambda_{ik}}\left[\log \pi_k(x_i;\nu) + \log \phi(y_i;h_k(x_i),\sigma_k^2)\right]}}
\end{align}
where
\begin{align}
    \lambda_{ik}
    &=\frac{p(y_i\mid Z_i=k,x_i;\gamma^\prime)\cdot P(Z_i=k\mid x_i;\gamma^\prime)}{\sum_{j=1}^{K}{p(y_i\mid Z_i=j,x_i;\gamma^\prime)\cdot P(Z_i=j\mid x_i;\gamma^\prime)}} \nonumber\\
    &= \frac{\phi(y_i;h_k(x_i),{\sigma_k^\prime}^2)\cdot \pi_k(x_i;\nu^\prime)}{\sum_{j=1}^{K}{\phi(y_i;h_j(x_i),{\sigma_j^\prime}^2)\cdot \pi_j(x_i;\nu^\prime)}}.\label{eq:lambda_ik}
\end{align}

Taking the derivative of $Q(\gamma\mid\gamma^\prime)$ w.r.t. $\sigma_j^2$ gives
\begin{align}
    \frac{\partial Q}{\partial\sigma_j^2} 
    &= \sum_{i=1}^{n}{\frac{\lambda_{ij}\cdot\frac{\partial}{\partial\sigma_j^2}\phi(y_i;h_j(x_i),\sigma_j^2)}{\phi(y_i;h_j(x_i),\sigma_j^2)}} \nonumber\\
    &=\sum_{i=1}^{n}{\frac{\lambda_{ij}[(h_j(x_i)-y_i)^2-\sigma_j^2]}{2\sigma_j^4}}.
\end{align}
Thus, 
\begin{equation}
    \widehat{\sigma}_j^2=\argmax_{\sigma_j^2}{Q(\gamma\mid\gamma^\prime)}=\frac{\sum_{i=1}^{n}{\lambda_{ij}[h_j(x_i)-y_i]^2}}{\sum_{i=1}^{n}{\lambda_{ij}}}, \quad j=1,\dots,K.
\end{equation}

There is no analytic expression for the optimal $\nu=\{\beta_1,\dots,\beta_K\}$ that maximizes $Q$. Thus 
\begin{equation*}
    \widehat{\nu}
    =\argmax_{\nu}{Q(\gamma\mid \gamma^\prime)}
    =\argmax_{\nu}{\sum_{i=1}^{n}{\sum_{k=1}^{K}{\lambda_{ik}\log \pi_k(x_i;\nu)}}}
\end{equation*} 
is found iteratively through the Fisher scoring algorithm given in Section~\ref{sec:project3_appendix_fisherScoring}.

\subsection{Fisher scoring update rule}\label{sec:project3_appendix_fisherScoring}
Here we derive the update rule for the $\nu$ parameter in the M-step of Algorithm~\ref{alg:project3_numerical_em}. 
\begin{equation*}
    \widehat{\nu}
    =\argmax_{\nu}{Q(\gamma\mid \gamma^\prime)}
    =\argmax_{\nu}{\sum_{i=1}^{n}{\sum_{k=1}^{K}{\lambda_{ik}\log \pi_k(x_i;\nu)}}}.
\end{equation*} 
When updating $\nu$, $\lambda_{ik}$ obtained from the E-step is considered fixed. 
Let $\lambda_i=(\lambda_{i1},\dots,\lambda_{iK})^\top$, $\pi_i=(\pi_{i1},\dots,\pi_{iK})^\top$ with
\begin{equation}
    \pi_{ik}=\pi_k(x_i;\nu)=\frac{\exp\{\beta_k\cdot\widehat{\mu}_k(x_i)\}}{\sum_{j=1}^{K}{\exp\{\beta_j\cdot\widehat{\mu}_j(x_i)\}}}.
\end{equation}
Then
\begin{align}
    l_i
    &= \sum_{k=1}^{K}{\lambda_{ik}\log \pi_{ik}} \\
    &=\sum_{k=1}^{K-1}{\lambda_{ik}\log\pi_{ik}+\left(1-\sum_{k=1}^{K-1}{\lambda_{ik}}\right)\log\left(1-\sum_{k=1}^{K-1}{\pi_{ik}}\right)}\\
    &= \sum_{k=1}^{K-1}{\lambda_{ik}\log\left(\frac{\pi_{ik}}{1-\sum_{j=1}^{K-1}{\pi_{ij}}}\right)}+\log\left(1-\sum_{k=1}^{K-1}{\pi_{ik}}\right)\\
    &= \sum_{k=1}^{K-1}{\lambda_{ik}\eta_{ik}}-A(\eta_i)
\end{align}
where $\eta_i=(\eta_{i1},\dots,\eta_{iK})^\top$ with
\begin{equation}
    \eta_{ik} = \log\left(\frac{\pi_{ik}}{1-\sum_{j=1}^{K-1}{\pi_{ij}}}\right)
    , \quad k=1,\dots,K
\end{equation}
and
\begin{equation}
    A(\eta_i)=-\log\left(1-\sum_{k=1}^{K-1}{\pi_{ik}}\right) = \log\left(\sum_{k=1}^{K}{e^{\eta_{ik}}}\right).
\end{equation}
We can write $\pi$ in terms of $\eta$
\begin{equation}
    \pi_{ik}=\frac{e^{\eta_{ik}}}{\sum_{j=1}^{K}{e^{\eta_{ij}}}}=\frac{\exp\{\beta_k\mu_{ik}\}}{\sum_{j=1}^{K}{\exp\{\beta_j\mu_{ij}\}}}, \quad k=1,\dots,K,
\end{equation}
where we write $\mu_{ik}=\widehat{\mu}_{k}(x_i)$ for simplicity. The second equality is defined by model specification $\eta_{ik}=\beta_k\mu_{ik}$ and $\beta_K=0$. So we have $K-1$ parameters to estimate $\beta_1,\dots,\beta_{K-1}$ since the $K$th term is used as reference level.

The log-likelihood of all data is
\begin{equation}
    l=\sum_{i=1}^{n}{l_i}=\sum_{i=1}^{n}{\left[-A(\eta_i)+\sum_{k=1}^{K-1}{\lambda_{ik}\eta_{ik}}\right]}.
\end{equation}
The score vector $u$ has elements 
\begin{equation}
    \frac{\partial l}{\partial\beta_s}=\sum_{i=1}^{n}{\sum_{k=1}^{K-1}{\frac{\partial l_i}{\partial\eta_{ik}}\cdot\frac{\partial \eta_{ik}}{\partial\beta_s}}}=\sum_{i=1}^{n}{(\lambda_{is}-\pi_{is})\mu_{is}}, \quad s=1,\dots,K-1,
\end{equation}
since 
\begin{equation}
    \frac{\partial A(\eta_i)}{\partial\eta_{ik}}=\pi_{ik},\quad \frac{\partial \eta_{ik}}{\partial\beta_s}=\mu_{ik}I(k=s), \quad k,s\in\{1,\dots,K-1\}.
\end{equation}
The Hessian matrix (in this case also the Fisher information) has elements
\begin{align}
    H_{st}=\frac{\partial ^2l}{\partial\beta_t\partial\beta_s}
    &=\frac{\partial}{\partial\beta_t}\left(\sum_{i=1}^{n}{(\lambda_{is}-\pi_{is})\mu_{is}}\right) \\
    &=\sum_{i=1}^{n}{-\mu_{is}\frac{\partial}{\partial\beta_t}\pi_{is}} \\
    &=\begin{cases}
    \sum_{i=1}^{n}{\mu_{is}\mu_{it}\pi_{is}\pi_{it}}, & s\neq t\\
    \sum_{i=1}^{n}{-\mu_{it}\mu_{it}\pi_{it}(1-\pi_{it})}, & s=t
    \end{cases}.
\end{align}

The score vector $u$ can be written in matrix form as $u=diag\{(\lambda-\pi)^T\mu\}$ where $\lambda$, $\pi$ and $\mu$ are $n\times (K-1)$ matrices with elements $\lambda_{ik}$, $\pi_{ik}$ and $\mu_{ik}$, respectively. $diag\{M\}$ extracts the main diagonal elements of $M$ as a column vector. $Diag\{M\}$ makes all elements not on the \textit{Main} diagonal of $M$ zero (i.e., return a diagonal matrix). $ODiag\{M\}$ makes all elements not on the \textit{Other} diagonal of $M$ zero. 

The Hessian matrix $H$ can be calculated in matrix notation as
\begin{equation}
    H=Diag\{(\mu\odot\pi)^\top[\mu\odot(1-\pi)]\} + ODiag\{(\mu\odot\pi)^\top(\mu\odot\pi)\}
\end{equation}
where $\odot$ is the Hadamard (element-wise) product of matrices.

\subsection{Algorithm~\ref{alg:project3_analytic_em} - EM with analytic updates}

In this section, we consider a different parametric form for the mixing coefficients which is given below. 
\begin{equation}
    \pi_k(x_i;\nu) = \frac{\beta_k \cdot p(\widehat{\mu}_k(x_i);\tau^2)}{p(x_i;\nu)}=\frac{\beta_k \cdot p(\widehat{\mu}_k(x_i);\tau^2)}{\sum_{j=1}^{K}{\beta_j\cdot p(\widehat{\mu}_j(x_i);\tau^2)}}
\end{equation}
with $\sum_{k=1}^{K}{\beta_k}=1$ and $\beta_k\ge0$. If $\widehat{\mu}_k(x)$ is estimating squared error, we further assume $p(\widehat{\mu}_k(x);\tau^2)$ is the density of a scaled chi-squared $\tau^2\chi^2_1$
\begin{equation}
    p(\widehat{\mu}_k(x);\tau^2)=\frac{1}{\tau^2\sqrt{2\pi}}\left(\frac{\widehat{\mu}_k(x)}{\tau^2}\right)^{-1/2}\exp\left\{-\frac{\widehat{\mu}_k(x)}{2\tau^2}\right\}.
\end{equation}

The complete-data likelihood is 
\begin{align}
    & \quad p(y,x,z;\gamma) \nonumber\\
    &= p(y\mid x,z;\gamma)\cdot p(z\mid x;\gamma)\cdot p(x;\gamma)\nonumber\\ 
    &= \prod_{i=1}^{n}{p(y_i\mid x_i,z_i;\gamma)\cdot p(z_i\mid x_i;\gamma)\cdot p(x_i;\gamma)} \nonumber\\
    &= \prod_{i=1}^{n}{\left[\prod_{k=1}^{K}{\phi(y_i;h_k(x_i),\sigma_k^2)^{I(z_i=k)}}\right] \cdot \left[\prod_{k=1}^{K}{\pi_k(x_i;\nu)^{I(z_i=k)}}\right] \cdot \left[\sum_{k=1}^{K}{\beta_k\cdot p(\widehat{\mu}_k(x_i);\tau^2)}\right]} \nonumber\\
    &=\prod_{i=1}^{n}{\left[\prod_{k=1}^{K}{\phi(y_i;h_k(x_i),\sigma_k^2)^{I(z_i=k)}}\right] \cdot \left[\prod_{k=1}^{K}{\{\beta_k\cdot p(\widehat{\mu}_k(x_i);\tau^2)\}^{I(z_i=k)}}\right]}.
\end{align}

The complete-data log-likelihood is thus
\begin{align}
    &\quad \log p(y,x,z;\gamma) \nonumber\\
    &= \sum_{i=1}^{n}{\sum_{k=1}^{K}{I(z_i=k)\left[\log\beta_k + \log p(\widehat{\mu}_k(x_i);\tau^2) + \log \phi(y_i;h_k(x_i),\sigma_k^2)\right]}}.
\end{align}

Marginalize out the hidden variables $Z$ by taking the expectation of the complete-data log-likelihood w.r.t. $p(z\mid y,x;\gamma^\prime)$ where $\gamma^\prime$ is some value of the parameter. This gives
\begin{align}
    &\quad Q(\gamma\mid\gamma^\prime)\nonumber\\
    &= \mathbb{E}\left[\log p(y,x,Z;\gamma)\mid y,x;\gamma^\prime\right] \nonumber\\
    &=\sum_{i=1}^{n}{\sum_{k=1}^{K}{\underbrace{P(Z_i=k\mid y_i,x_i;\gamma^\prime)}_{\lambda_{ik}}\left[\log\beta_k + \log p(\widehat{\mu}_k(x_i);\tau^2) + \log \phi(y_i;h_k(x_i),\sigma_k^2)\right]}}
\end{align}
where
\begin{align}
    \lambda_{ik}
    &=\frac{p(y_i\mid Z_i=k,x_i;\gamma^\prime)\cdot P(Z_i=k\mid x_i;\gamma^\prime)}{\sum_{j=1}^{K}{p(y_i\mid Z_i=j,x_i;\gamma^\prime)\cdot P(Z_i=j\mid x_i;\gamma^\prime)}} \nonumber\\
    &= \frac{\phi(y_i;h_k(x_i),{\sigma_k^\prime}^2)\cdot \pi_k(x_i;\nu^\prime)}{\sum_{j=1}^{K}{\phi(y_i;h_j(x_i),{\sigma_j^\prime}^2)\cdot \pi_j(x_i;\nu^\prime)}}.\label{eq:lambda_ik_analytic_em}
\end{align}

Taking the derivative of $Q(\gamma\mid\gamma^\prime)$ w.r.t. $\sigma_j^2$ gives
\begin{align}
    \frac{\partial Q}{\partial\sigma_j^2} 
    &= \sum_{i=1}^{n}{\frac{\lambda_{ij}\cdot\frac{\partial}{\partial\sigma_j^2}\phi(y_i;h_j(x_i),\sigma_j^2)}{\phi(y_i;\mu_j(x_i),\sigma_j^2)}} \nonumber\\
    &=\sum_{i=1}^{n}{\frac{\lambda_{ij}[(h_j(x_i)-y_i)^2-\sigma_j^2]}{2\sigma_j^4}}.
\end{align}
Thus, 
\begin{equation}
    \widehat{\sigma}_j^2=\argmax_{\sigma_j^2}{Q(\gamma\mid\gamma^\prime)}=\frac{\sum_{i=1}^{n}{\lambda_{ij}[h_j(x_i)-y_i]^2}}{\sum_{i=1}^{n}{\lambda_{ij}}}, \quad j=1,\dots,K.
\end{equation}

To find $\beta=\{\beta_1,\dots,\beta_K\}$ that maximizes $Q(\gamma\mid \gamma^\prime)$ with $\sum_{k=1}^{K}{\beta_k}=1$, we use Lagrange multiplier which gives
\begin{equation}
    \widehat{\beta}_j=\frac{1}{n}\sum_{i=1}^{n}{\lambda_{ij}}, \quad j=1,\dots,K.
\end{equation}

To find $\tau^2$ that maximizes $Q(\gamma\mid\gamma^\prime)$, we have
\begin{align}
    \frac{\partial Q}{\partial\tau^2} 
    &= \sum_{i=1}^{n}{\sum_{k=1}^{K}{\lambda_{ik}\cdot\frac{\partial }{\partial\tau^2}\log p(\widehat{\mu}_k(x_i);\tau^2) }} \nonumber \\
    &= \sum_{i=1}^{n}{\sum_{k=1}^{K}{\lambda_{ik}\cdot\frac{\widehat{\mu}_k(x_i)-\tau^2}{2\tau^4} }}.
\end{align}
Thus, 
\begin{equation}
    \widehat{\tau}^2=\argmax_{\tau^2}{Q(\gamma\mid\gamma^\prime)}=\frac{1}{n}\sum_{i=1}^{n}{\sum_{k=1}^{K}{\lambda_{ik}\cdot \widehat{\mu}_k(x_i)}}.
\end{equation}

\begin{algorithm}
\caption{Model combination with analytic EM}\label{alg:project3_analytic_em}
\nl Initialize parameter estimates $\gamma^{(0)}$\;
\For{t=1,2,\dots}{
\begin{enumerate}
    \item E-step: calculate 
    \begin{equation*}
        \lambda_{ik}=\frac{\phi\left(y_i;h_k(x_i),{\sigma_k^2}^{(t-1)}\right)\cdot \pi_k(x_i;\nu^{(t-1)})}{\sum_{j=1}^{K}{\phi\left(y_i;h_j(x_i),{\sigma_j^2}^{(t-1)}\right)\cdot \pi_j(x_i;\nu^{(t-1)})}}, \quad i=1,\dots,n, \quad k=1,\dots,K
    \end{equation*}
    \item M-step: 
    \begin{align*}
        {\sigma_j^2}^{(t)} 
        &= \frac{\sum_{i=1}^{n}{\lambda_{ij}[h_j(x_i)-y_i]^2}}{\sum_{i=1}^{n}{\lambda_{ij}}}, \quad j=1,\dots,K \\
        \beta_j^{(t)} &= \frac{1}{n}\sum_{i=1}^{n}{\lambda_{ij}}, \quad j=1,\dots,K \\
        {\tau^2}^{(t)} &= \frac{1}{n}\sum_{i=1}^{n}{\sum_{k=1}^{K}{\lambda_{ik}\cdot \widehat{\mu}_k(x_i)}}
    \end{align*}
\end{enumerate}
}
\end{algorithm}

\section{Illustration of model combination using PASD ensembles with two classification examples}\label{sec:project3_appendix_illustration_of_mc_classification}

\begin{figure}[!htb]
    \centering
    \includegraphics[scale=0.7]{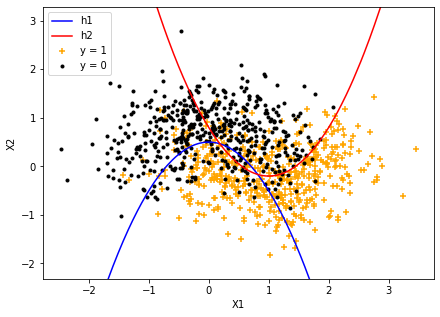}
    \caption{Visualization for the proposed model combination method in a one-dimensional classification setting. The models $h_1$ and $h_2$ are the individual prediction models for predicting the binary outcome $Y$ given $X$. Both models only perform well in certain parts of the covariate space.}
    \label{fig:mv_illustration_classfication_makeMoons_dataAndmodel}
\end{figure}

For illustrations in classification problems, we consider two synthetic datasets from scikit-learn \cite{Pedregosa2011Scikit-learn:Python}. The first one is the \textit{make\_moons} dataset which has two covariates $X_1$ and $X_2$, and a binary outcome $Y\in\{0,1\}$. If we take a sample $D$ of size 1000 and plot both covariates and represent the outcome $Y$ with different markers (as in Figure~\ref{fig:mv_illustration_classfication_makeMoons_dataAndmodel}), then the plot looks like two interleaving half circles. 

\begin{figure}[!htb]
    \centering
    \includegraphics[scale=0.7]{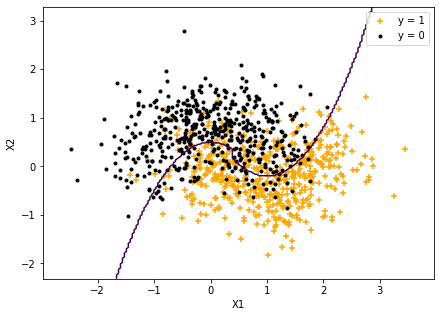}
    \caption{Visualization for the proposed model combination method in a one-dimensional classification setting. The decision boundary of the combined model for predicting the binary outcome $Y$ given $X$ is presented, which shows a much better fit to the data than any of the two individual models.}
    \label{fig:mv_illustration_classification_makeMoons_combined_model}
\end{figure}

We consider two predictions model $h_1(x)=I\{0.5-X_1^2-X_2>0\}$ and $h_2(x)=I\{(X_1-1)^2-0.2-X_2>0\}$. Their decision boundaries are plotted in Figure~\ref{fig:mv_illustration_classfication_makeMoons_dataAndmodel}. Both models only do well in certain parts of the covariate space and by applying the Majority voting guided model combination algorithm (Algorithm~\ref{alg:project3_mv_combination}), we increase the classification accuracy (proportion of observations that are correctly classified) on data $D$ of the two individual models, $0.62$ and $0.63$, to a combined model accuracy of $0.81$. The decision boundary of the combined model is shown in Figure~\ref{fig:mv_illustration_classification_makeMoons_combined_model} where we see it learns to give higher weights to models where they are estimated to do well and avoid parts where they are estimated to perform poorly.

\begin{figure}[!htb]
    \centering
    \includegraphics[scale=0.7]{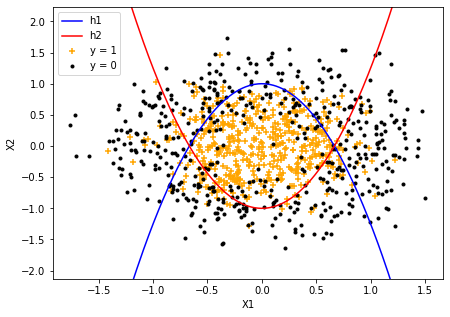}
    \caption{A second visualization for the proposed model combination method in a one-dimensional classification setting. The models $h_1$ and $h_2$ are the individual prediction models for predicting the binary outcome $Y$ given $X$. Again, both models only perform well in certain parts of the covariate space.}
    \label{fig:mv_illustration_classfication_makeCircles_dataAndmodel}
\end{figure}

We illustrate the algorithm with a second example of a classification task that uses the dataset \textit{make\_circles}. Again it is a dataset with binary outcome $Y\in\{0,1\}$ with two covariates $X_1$ and $X_2$ where the plot looks like a large circle ($Y=0$ cases) containing a small circle ($Y=1$ cases) as shown in Figure~\ref{fig:mv_illustration_classfication_makeCircles_dataAndmodel}.

\begin{figure}[!htb]
    \centering
    \includegraphics[scale=0.7]{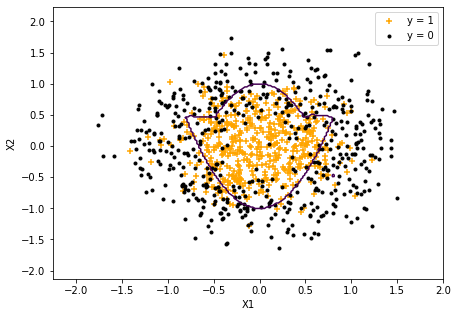}
    \caption{The second visualization for the proposed model combination method in a one-dimensional classification setting. The decision boundary of the combined model for predicting the binary outcome $Y$ given $X$ is presented, which again shows a much better fit to the data than any of the two individual models.}
    \label{fig:mv_illustration_classification_makeCircles_combined_model}
\end{figure}

We consider two prediction models $h_1(x)=I\{1-(1.5X_1)^2-X_2>0\}$ and $h_2(x)=I\{(1.5X_1)^2-1-X_2<0\}$ with classification accuracy $0.7$ and $0.72$, respectively. Their decision boundaries are plotted in Figure~\ref{fig:mv_illustration_classfication_makeCircles_dataAndmodel}. If we apply Algorithm~\ref{alg:project3_mv_combination}, we obtain a combined model with classification accuracy $0.79$ with decision boundary shown in Figure~\ref{fig:mv_illustration_classification_makeCircles_combined_model}. The combined prediction model is able to discover the correct regions of the covariate space to combine the two individual models. 

\section{Additional simulation results for EM-based model combination}

In this section, we used simulations to evaluate the performance of the proposed EM-based model combination algorithm. We simulated a 10-dimensional covariate vector $(X_1,\dots,X_{10})$, where each $X_i, i \in \{1, \ldots, 10\}$ are iid $Uni(0,1)$. The outcome was simulated using
\begin{equation}
    Y=10\sin(\pi X_1X_2)+20(X_3-0.5)^2+10X_4+5X_5+\epsilon 
\end{equation}
where $\epsilon\sim N(0,2X_4)$. Only five of the covariates are related to the outcome and the remaining ones are independent of $Y$. This example is taken from \cite{Friedman1991MultivariateSplines}, but we modified the error distribution to make it dependent on one of the covariates.

\begin{figure}[!htbp]
    \centering
    \begin{subfigure}[b]{0.45\textwidth}
         \centering
         \includegraphics[width=\textwidth]{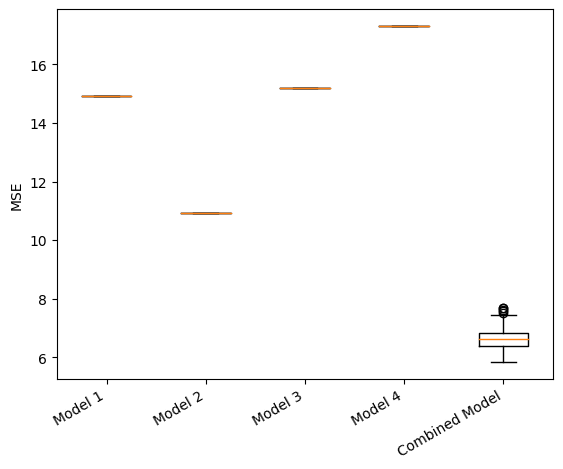}
         \caption{MSE}
         \label{fig:mse_emModelComb_friedman1}
     \end{subfigure}
     \qquad
     \begin{subfigure}[b]{0.45\textwidth}
         \centering
         \includegraphics[width=\textwidth]{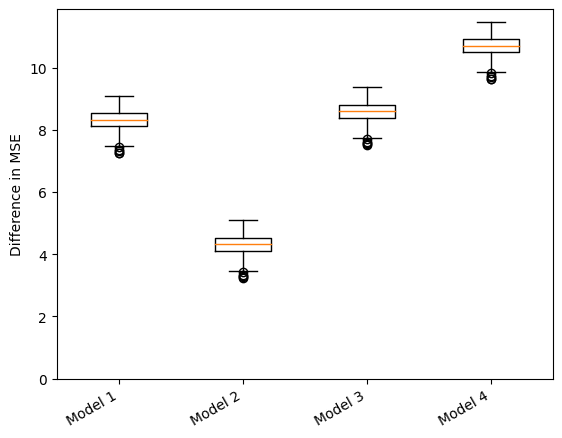}
         \caption{Improvement in MSE}
         \label{fig:relative_mse_emModelComb_friedman1}
     \end{subfigure}
    \caption{Comparison between combined model and individual models for the simulation results of EM-based model combination. The combined model has lower MSE on the test dataset than all four individual prediction models in all 1000 simulation runs.}
    \label{fig:emModelComb_friedman1}
\end{figure}

We considered four given elastic net models for predicting the outcome $Y$ from $X$. Each of the four models missed certain covariates that are related to the outcome. We randomly sampled a dataset $D_{combine}$ of size 1000 from the true data generating process, and it was used in the EM-based model combination algorithm to estimate the parameters of the weights. This process was repeated 1000 times and the 1000 final combined models were then evaluated on the same held-out test dataset $D_{test}$ of size 10000. The results are presented in Figure~\ref{fig:emModelComb_friedman1}.

Figure~\ref{fig:mse_sglvsEnsem_Friedman1_lm} shows the distributions of the MSE of four individual models and the combined model when evaluated on the test dataset in all 1000 simulation runs. The combined model always had lower MSE compared to the four individual prediction models. Figure~\ref{fig:relative_mse_emModelComb_friedman1} looks at the MSE difference by subtracting the MSE of the combined model. That is, in each of the 1000 simulations of the experiment, we calculate the MSE for each of the four individual prediction models and subtract the combined model MSE from it. It is clear that the combined model always had lower MSE (i.e., were always better) than the individual models in all of the 1000 simulations as indicated by the positive differences.

\section{Additional information on the four lung cancer risk models}\label{sec:project3_appendix_lcModels_additional_info}

In this section, we list the risk factors used in each of the four lung cancer risk models in Table~\ref{tab:project3_four_model_covariates}.

\begin{table}[!htb]
\centering
\caption{List of risk factors considered in each lung cancer risk prediction model.}
\label{tab:project3_four_model_covariates}
\resizebox{0.9\textwidth}{!}{%
\begin{tabular}{@{}ccccc@{}}
\toprule
Risk factors                  & Bach & PLCOm2012 & PLCOm2012 (no race) & PLCOm2012 (simplified) \\ \midrule
Age                           & Y    & Y         & Y                 & Y                     \\
Gender                        & Y    & -         & -                 & -                     \\
Education                     & -    & Y         & Y                 & -                     \\
Race/Ethnicity                & -    & Y         & -                 & -                     \\
BMI                           & -    & Y         & Y                 & -                     \\
COPD                          & -    & Y         & Y                 & -                     \\
Asbestos exposure             & Y    & -         & -                 & -                     \\
Personal history of cancer    & -    & Y         & Y                 & -                     \\
Family history of lung cancer & -    & Y         & Y                 & -                     \\
Smoking status                & -    & Y         & Y                 & Y                     \\
Cigarettes per day            & Y    & Y         & Y                 & Y                     \\
Smoking duration              & Y    & Y         & Y                 & Y                     \\
Quit duration                 & Y    & Y         & Y                 & Y                     \\ \bottomrule
\end{tabular}%
}
\end{table}

We also provide the sample correlation between any pair of the four lung cancer risk models in Table~\ref{tab:project3_corr_M1_M4} calculated using the NLST data

\begin{table}[]
\centering
\caption{Sample correlations between predicted risks on $D_2$.}
\label{tab:project3_corr_M1_M4}
\resizebox{\textwidth}{!}{%
\begin{tabular}{c|cccc}
\hline
                       & Bach  & PLCOm2012 & PLCOm2012 (no race) & PLCOm2012 (simplified) \\ \hline
Bach                   & 1.000 & 0.793     & 0.815               & 0.933                  \\
PLCOm2012              &       & 1.000     & 0.977               & 0.843                  \\
PLCOm2012 (no race)    &       &           & 1.000               & 0.866                  \\
PLCOm2012 (simplified) &       &           &                     & 1.000                  \\ \hline
\end{tabular}%
}
\end{table}

\subsection{Comparison of single PASD trees and PASD ensembles}
\label{app:nlst-ensemble}
\begin{figure}[!htbp]
    \centering
    \begin{subfigure}[b]{0.45\textwidth}
         \centering
         \includegraphics[width=\textwidth]{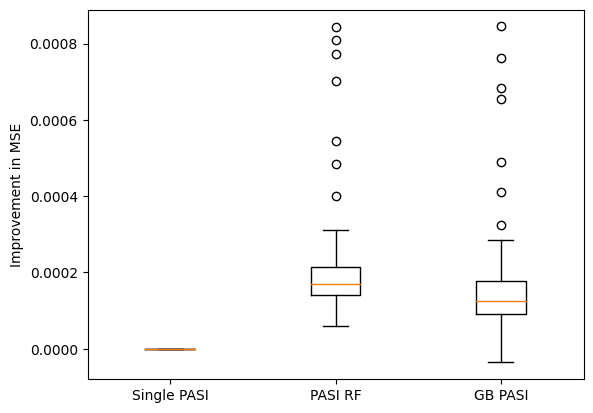}
         \caption{Bach}
         \label{fig:relative_mse_sglvsEnsem_NLST_bach}
     \end{subfigure}
     \qquad
     \begin{subfigure}[b]{0.45\textwidth}
         \centering
         \includegraphics[width=\textwidth]{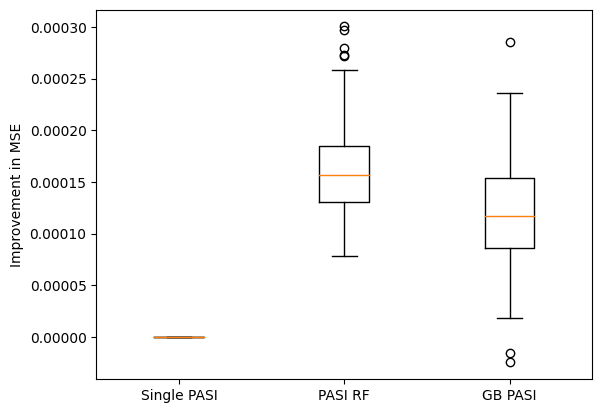}
         \caption{PLCOm2012}
         \label{fig:relative_mse_sglvsEnsem_NLST_plco}
     \end{subfigure}
     \begin{subfigure}[b]{0.45\textwidth}
         \centering
         \includegraphics[width=\textwidth]{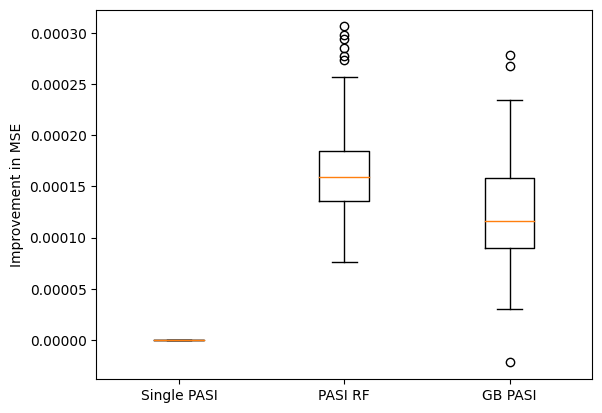}
         \caption{PLCOm2012 (no race)}
         \label{fig:relative_mse_sglvsEnsem_NLST_plcoNoRace}
     \end{subfigure}
     \qquad
     \begin{subfigure}[b]{0.45\textwidth}
         \centering
         \includegraphics[width=\textwidth]{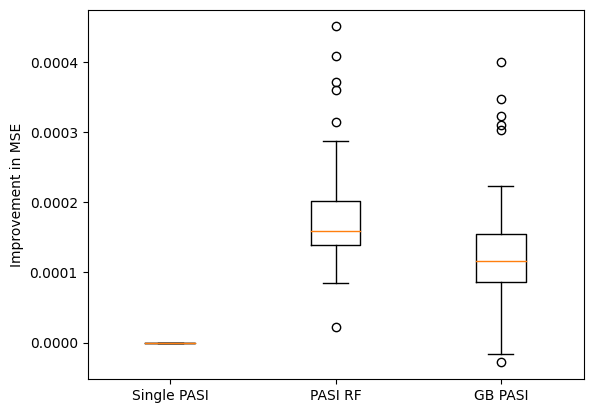}
         \caption{PLCOm2012 (simplified)}
         \label{fig:relative_mse_sglvsEnsem_NLST_plcoSimp}
     \end{subfigure}
    \caption{Improvements in MSE of random PASD forest and gradient boosting PASD trees (relative to a single PASD tree) when they are used to estimate the Brier scores of the four risk models: Bach, PLCOm2012, PLCOm2012 without race, PLCOm2012 simplified.}
    \label{fig:relative_mse_sglvsEnsem_NLST}
\end{figure}

To compare the performance of the single PASD trees and the ensemble of PASD trees we estimate the Brier scores of the four models M1-M4 when applied to the NLST data. We divided the NLST data into a training and a test set with sample sizes of $9781$ and $39,128$, respectively. For each of the four risk models, we fit the single PASD tree and the PASD ensembles on the training set and evaluated their MSE for estimating the Brier score using the test set. We repeated the process for 100 different splits into a training and a test set and the MSE of the single PASD tree and the PASD ensembles for each of the four models are across the 100 splits are shown in boxplots  in Figure~\ref{fig:relative_mse_sglvsEnsem_NLST}. In all four cases and for all splits, the tree ensembles had smaller MSE than the single PASD tree.

\section{Application of PASD trees to the COMPAS Data}\label{sec:project2_compas_analysis}

In this section, we applied the PASD tree algorithm to the Correctional Offender Management Profiling for Alternative Sanctions (COMPAS) risk scores system \cite{Corbett-Davies2016AClear.} which is used by many courts in the US to assess the likelihood of defendant recidivism. There has been debate over whether the risk score prediction is associated with potential racial and/or gender bias \cite{Corbett-Davies2016AClear.,Angwin2016MachineBias}. However, our analysis was not focused on that particular question. As a data-driven approach, the PASD tree tries to discover potential subgroups from data (rather than pre-specifying them) that may have differential model performance under the COMPAS risk score. 

We used the same dataset as in the \cite{Angwin2016MachineBias} analysis. It contains, for each defendant, their demographic information (age, sex, race), crime records (degree of charge, number of priors), the COMPAS risk score (1-10) and a binary indicator of whether the person is arrested again within two years. We considered specificity and sensitivity as the model performance measures; and like in the NLST example, we split the data into training (for fitting PASD trees) and test (for evaluating fitted PASD trees). We also used 1000 seeds for cross-validation and looked at the trees that were most frequently selected. 

\begin{figure}[h!]
\centering
\includegraphics[width=\textwidth]{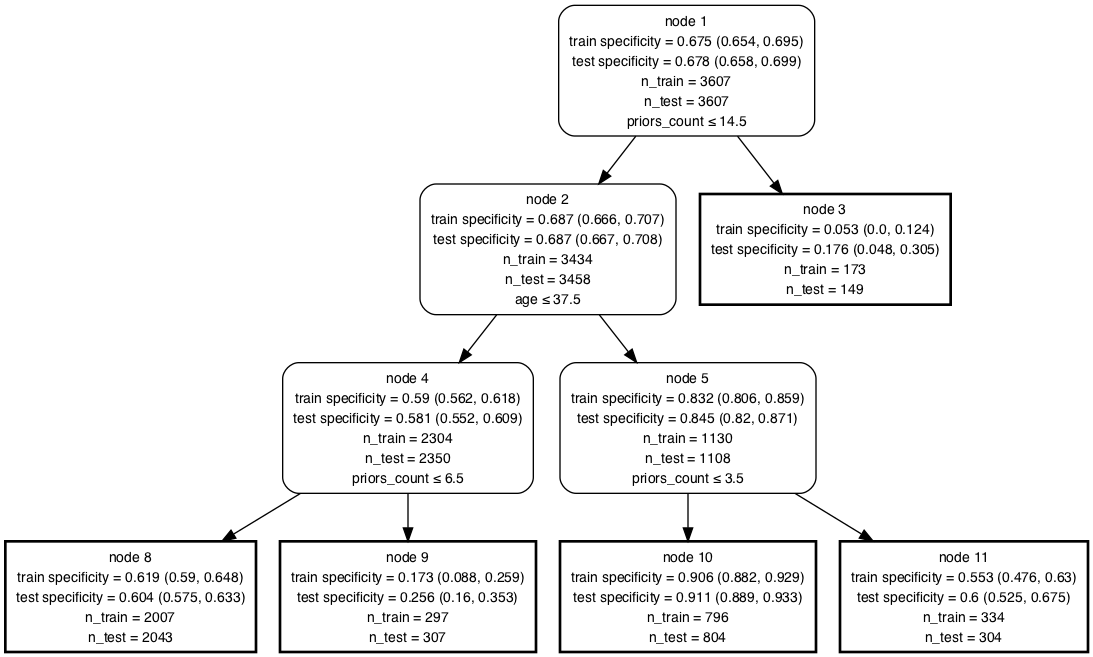}
\caption{Most frequently selected tree by the PASD algorithm applied to the COMPAS data with specificity as the model performance measure.}
\label{compas_spe_PASD_most_selected_tree}
\end{figure}

\begin{figure}[h!]
\centering
\includegraphics[scale=0.4]{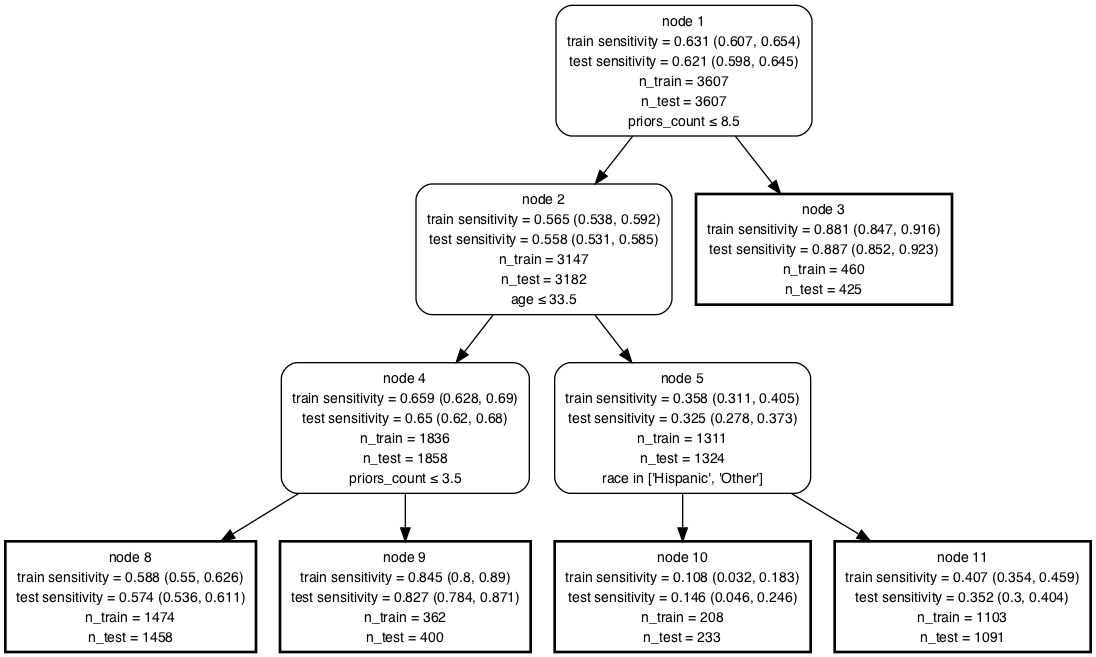}
\caption{Most frequently selected tree by the PASD algorithm applied to the COMPAS data with sensitivity as the model performance measure.}
\label{compas_sen_PASD_most_selected_tree}
\end{figure}

The results are given in Figure~\ref{compas_spe_PASD_most_selected_tree} and \ref{compas_sen_PASD_most_selected_tree}. When specificity is used as performance measure, the final selected tree splits only on \textit{priors\_count} and \textit{age}. Looking at the final subgroups, we see the risk scores have low specificity (less than $0.3$) for people with either \{$priors\_count>14.5$\} or \{$6.5<priors\_count\le14.5$ and $age\le37.5$\}, medium specificity (around $0.6$) for subgroups \{$priors\_count\le6.5$ and $age\le37.5$\} and \{$3.5<priors\_count\le14.5$ and $age>37.5$\}, and finally high specificity (around $0.9$) for people with \{$priors\_count\le3.5$ and $age>37.5$\}. Potential subgroups with differential sensitivity under the risk scores can be read from Figure~\ref{compas_sen_PASD_most_selected_tree} in a similar manner.

\end{document}